\documentclass[11pt]{article}

\usepackage{lineno}
\usepackage[utf8]{inputenc}
\usepackage{bm}
\usepackage{amsmath}
\usepackage{enumerate}
\usepackage{xcolor}
\usepackage{graphicx}
\usepackage{subcaption}
\usepackage{float} 
\usepackage{amssymb}
\usepackage{soul}
\usepackage{longtable}
\usepackage{mathtools}
\usepackage{amsmath}
\usepackage[margin=0.75in]{geometry}
\usepackage{hyperref}
\usepackage[round]{natbib}
\usepackage{authblk}

\definecolor{nika}{rgb}{0.5,0,0.5}

\definecolor{anur}{rgb}{0.0,0.0,1.0}

\title{Normalized Field Product method for Topology Optimization}
\author[1]{Nikhil Singh}
\author[2]{Anupam Saxena}
\affil[1]{Indian Institute of Technology Kanpur, 208016, India\\ singhn@iitk.ac.in}
\affil[2]{Indian Institute of Technology Kanpur, 208016, India\\ anupams@iitk.ac.in}
\date{\today}

\numberwithin{equation}{section}

\begin{document}
	\maketitle
	\begin{abstract}
		The paper presents a novel, parameter free, density evaluation method for topology optimization based on normalized product of a scalar field. The approach imposes length scale on solid phase implicitly and allows for pure 0-1 singularity free solutions to exist within the design space. The density formulation proposed herein is independent of user specified parameters, other then the length scale. Sensitivity analysis for the formulation obtained by combining the proposed density evaluation with the SIMP material model reveal that for compliance minimization and small deformation compliant mechanism problems, objective gradients vanish at pure 0-1 topologies. The methodology is implemented to solve well known examples in literature for different mesh refinements presenting mesh independent and close to binary solutions without the use of heuristics or continuation methods.
	\end{abstract}
	\section{Introduction}
\label{sec:Intro}
%	\begin{enumerate}
%		\item Description of a typical density TO problem 
%		\item Checker board pattern
%		\item Filtering and its associated proofs
%		\item Projection --- close to black and white solutions
%	\end{enumerate}
Since its inception \cite{Bendsoe1989}, topology optimization is often expressed as a material distribution problem that of determining the optimal distribution of material within a design domain, $\Omega$, to minimize a specified objective. In density method, presence or absence of material at a point is expressed using an indicator \textit{density field}, $\rho(\boldsymbol{\mathrm{X}})$. The aim is to determine the density distribution, $0 \leq \rho(\boldsymbol{\mathrm{X}}) \leq 1$ for $\boldsymbol{\mathrm{X}} \in \Omega$, where $\rho(\boldsymbol{\mathrm{X}}) = 1$ and $\rho(\boldsymbol{\mathrm{X}}) = 0$ imply presence and absence of material at $\boldsymbol{\mathrm{X}}$ respectively. It is well known that the topology optimization problem in its original continuum form is ill-posed \citep{Sigmund1998,allaire1993numerical} because it lacks closure of the solution space and thus, is deficient of existence of solution. This lack of existence of solution is reflected through mesh dependent solutions in the numerical framework. Numerical problems are however, closed and have existence of solution as the minimum feature size is governed by the size of element. Proving existence of solution for the continuum formulation requires an extensive/rigorous mathematical proof. Most works however, resort to presenting mesh independent solutions for the corresponding numerical problem which indicates existence of solution but does not prove it. A common approach to ensure existence of solution is to restrict the solution space \citep{Eschenauer2001}. Over years, multiple approaches have been proposed. \cite{ambrosio1993optimal} proposed the perimeter constraint method which was numerically implemented by \cite{Haber1996}. The perimeter constraint leads to mesh independent solutions but choosing the appropriate bound on the perimeter requires some level of experience. Additionally, the method may lead to infeasible solutions as it does not prevent formation of thin members.

Another, readily implemented approach for imposing restriction is disallowing rapid variations in the density field. \cite{Petersson1998} introduce local constraint on density gradient thereby capping the maximum density variation between adjacent elements. \cite{sigmund200199} and \cite{sigmund1997design} implement a sensitivity filter which modifies the sensitivity heuristically. Local density gradient constraint and sensitivity filters lead to similar mesh independent results. An issue with sensitivity filters is that as the modifications are based on heuristics, it is unclear as to which mathematical problem is being solved. \cite{bruns2001topology} introduce density filtering which expresses the density of an element as the weighted sum of element densities of the neighboring elements thereby introducing a sense of smoothness to density evaluations. Weights in \cite{bruns2001topology} are calculated using a linear decaying function while \cite{bruns2003element} and \cite{wang2005bilateral} evaluate weights using a Gaussian distribution function. Density filters are known to produce mesh independent solutions. \cite{bourdin2001filters} prove existence of solution for a general case of density filtering. The drawback of these approaches is that they result in gray regions at the solid-void interface which is undesirable. Widths of these regions are dependent on the filtering radius and the choice of weight function. Ideally, a solution to the topology optimization problem should have no such transition region and be purely 0-1.

An alternate approach of dealing with rapid density variations is to introduce length scales to the problem formulation. \cite{Poulsen2003} implemented a local length scale constraint which imposed the length scale by capping the number of phase changes within a circular region around a point, referred to as the \textit{looking glass}. This method leads to a large number of constraints. To overcome this issue, multiple global length scale constraints have been proposed \citep{Zhang2014,Guo2014,Xia2015,Singh2020}. Imposition of an explicit minimum length scale ensures existence of solution but is known to hinder the optimization process \citep{Guo2014,Allaire2016}. \cite{Singh2020} discuss a heuristic methodology to prevent convergence to undesired local minima when imposing length scale constraints. \cite{Guest2004} proposes a projection scheme which leads to an implicit imposition of the length scale and provides close to black and white solutions. \cite{Guest2004} present mesh independent solutions for the compliance minimization problem. Projection introduces a parameter which provides implicit control over the width of a transition region. Thus, a projection method can produce close to 0-1 solutions but a pure 0-1 solution is not part of the solution space. \cite{Guest2009a} extends the projection method to impose length scale on both solid and void phases. A more detailed discussion on density filtering and projection is provided in section \ref{sec:formulation_def}. Besides projection, \cite{sigmund2007morphology} introduces morphological filters which also produce close to 0-1 solutions. A drawback of these approaches is that they rely on user choices of functions and parameters. Filtering requires a choice of weight function, while a proper implementation of the projection method and morphological filters requires both, a choice of weight function and implementing continuation on system parameters. Over the years, works have experimented with various weight functions but have not reported any significant advantage of implementing any specific weight function \citep{bruns2003element,wang2005bilateral}. Also, no standard method for applying continuation on system parameters has been established. Ideally, a topology optimization method should be void of any heuristics.

The approach presented in this paper allows for pure 0-1 solutions, imposes length scales implicitly and is free of any user specified parameters and heuristics. The only parameter specified corresponds to the length scale. We present a novel density evaluation method based on product of a scalar field over a domain defined in section \ref{sec:Math}. Unlike filtering and projection, the proposed density evaluation method includes a pure 0-1 solution in the solution space. Further, we implement the simplified isotropic material with penalization (SIMP) material model to evaluate stiffness for intermediate densities. The topology optimization formulation herein is shown to produce mesh independent, pure 0-1 solutions without additional constraints/filters. 
%Additionally, it is shown that pure 0-1 solutions are extrema for both compliance minimization and compliant mechanism problems.
%

Rest of the paper is organized as follows: section \ref{sec:Math} presents some mathematical pre-requisites that include defining, evaluating and investigating properties of product of a scalar field over a domain. Section \ref{sec:formulation_def} establishes some criteria to identify an ideal topology optimization formulation and discusses known methods under these criteria, followed by introduction of an alternate approach. Section \ref{sec:FPM} presents a novel density evaluation formulation based on product of a field established in section \ref{sec:Math}, followed by sensitivity analysis required for implementing gradient based optimization. Analysis in section \ref{sec:Math} is prompted by discussions in section \ref{sec:formulation_def} but is presented beforehand to maintain continuity between section \ref{sec:formulation_def} and \ref{sec:FPM}. Thus, one may consider directly perusing through section \ref{sec:formulation_def} and reverting to section \ref{sec:Math} when one feels. Section \ref{sec:example} presents problem description and solutions to some well known problems in topology optimization obtained using the formulation developed in section \ref{sec:FPM}. This is followed by discussion in section \ref{sec:discussion} and conclusions in section \ref{sec:conclusion}.

%	\pagebreak
%
	\section{Mathematical Pre-requisites}
	\label{sec:Math}
	This section establishes mathematical basis to construct the desired density evaluation method for the topology optimization formulation in section \ref{sec:FPM}.  
%	Definitions and symbols in this section are independent from those in the previous section (section \ref{sec:formulation_def}). 
	A definition for the product of a scalar field is proposed in section \ref{sec:PF}. Analytical expression for its evaluation is established thereafter, followed by discussion on one of its properties in section \ref{sec:conv_zero}. A method for numerical evaluation of the aforementioned analytical expression is developed in section \ref{sec:Num_eval}.
	\subsection{Product of a field}
	\label{sec:PF}
	Let $\gamma(\boldsymbol{\mathrm{x}}) > 0, \boldsymbol{\mathrm{x}} \in \Omega$ be a scalar field defined over a region $\Omega$ and $P = \{\Omega_1, \Omega_2,\hdots, \Omega_n\}$ be the partition of $\Omega$. Then, the product, $G$, of $\gamma(\boldsymbol{\mathrm{x}})$ over $\Omega$ is defined as
	\begin{eqnarray}
	G = \lim\limits_{A_i \to 0}~ \prod_{i = 1}^{n} \gamma(\boldsymbol{\mathrm{x}}_i)^{A_i}.
	\label{eqn:Prod_def}
	\end{eqnarray}
	where $\boldsymbol{\mathrm{x}}_i$ and $A_i$ are the centroid and area/volume of $\Omega_i$ respectively. The above definition is very similar to the Riemann sum used to define integrals. The limit $A_i \to 0$ implies $n \to \infty$ because, as the area of each partition becomes infinitesimally small, infinitely many partitions are required to cover $\Omega$. Also, as $A_i \to 0$, partition $\Omega_i$ converges to $\boldsymbol{\mathrm{x}}_i$.
	\subsubsection{Evaluation of $G$}
	\label{sec:Limit_eval}
	An analytical expression for evaluation of the limit in eqn. \ref{eqn:Prod_def} is presented. Taking natural logarithm on both sides of eqn. \ref{eqn:Prod_def} gives,
	\begin{eqnarray}\nonumber
	\ln G &=& \lim\limits_{ A_i \to 0}~ \sum\limits_{i=1}^{n} \ln (\gamma_i) A_i \\ \nonumber
	&=& \int\limits_{\Omega} \ln \gamma(\boldsymbol{\mathrm{x}}) d\Omega \\ 
	\implies G &=& \exp \left(\int\limits_{\Omega} \ln \gamma(\boldsymbol{\mathrm{x}}) d\Omega \right).
	\label{eqn:Prod_analytical}
	\end{eqnarray}
	Thus, the limit in eqn. \ref{eqn:Prod_def} exists so long as the integral in eqn. \ref{eqn:Prod_analytical} is defined. The analytical expression in eqn. \ref{eqn:Prod_analytical} is referred to as the \textit{Field Product} or FP formula. A corollary of the above result is
	\begin{eqnarray}
	G_n = \lim\limits_{ A_i \to 0}~ \prod_{i = 1}^{n} \gamma(\boldsymbol{\mathrm{x}}_i)^{A_i/A(\Omega)} = \exp \left( \dfrac{\displaystyle\int\limits_{\Omega} \ln \gamma(\boldsymbol{\mathrm{x}}) d\Omega}{\displaystyle\int\limits_{\Omega} d\Omega} \right)
	\label{eqn:Field_prod}
	\end{eqnarray} 
	where $A(\Omega)$ is the area or volume of $\Omega$. Eqn. \ref{eqn:Field_prod} is obtained by replacing $A_i$ with $A_i/A(\Omega)$ in eqn. \ref{eqn:Prod_def} and following the same evaluation procedure implemented to obtain eqn. \ref{eqn:Prod_analytical}. Replacing $A_i$ by $A_i/A(\Omega)$ non-dimensionalizes the exponent. The expression in eqn. \ref{eqn:Field_prod} is essential to the new parameter free density formulation discussed later and is referred to as \textit{Normalized Field Product} or nFP formula. We next investigate the limiting case when $\gamma(\boldsymbol{\mathrm{x}}) \to 0$ within a region of the domain.
	\subsubsection{Limiting case $\gamma(\boldsymbol{\mathrm{x}}) \to 0$}
	\label{sec:conv_zero}
	If $\gamma(\boldsymbol{\mathrm{x}}) \to 0$ for $\boldsymbol{\mathrm{x}} \in \mathcal{B}$ where $\mathcal{B} \subset \Omega$ is a finite region and $\gamma(\boldsymbol{\mathrm{x}})$ takes finite values for $\boldsymbol{\mathrm{x}} \in \Omega - \mathcal{B}$, then nFP of $\gamma(\boldsymbol{\mathrm{x}})$ goes to 0. To show this, we split the expression in eqn. \ref{eqn:Field_prod} as follows,
	\begin{eqnarray}
	G_n = \exp \left( \dfrac{\displaystyle\int\limits_{\Omega} \ln \gamma(\boldsymbol{\mathrm{x}}) d\Omega}{\displaystyle\int\limits_{\Omega} d\Omega} \right)
	&=& \underbrace{\exp \left( \dfrac{\displaystyle\int\limits_{\Omega - \mathcal{B}} \ln \gamma(\boldsymbol{\mathrm{x}}) d\Omega}{\displaystyle\int\limits_{\Omega} d\Omega} \right)}_\text{Term 1} \times \underbrace{\exp \left( \dfrac{\displaystyle\int\limits_{\mathcal{B}} \ln \gamma(\boldsymbol{\mathrm{x}}) d\Omega}{\displaystyle\int\limits_{\Omega} d\Omega} \right)}_\text{Term 2}.
	\label{eqn:conv_to_zero}
	\end{eqnarray}
	As $\gamma(\boldsymbol{\mathrm{x}})$ for $\boldsymbol{\mathrm{x}} \in \Omega - \mathcal{B}$ is finite, Term 1 in the expression above is finite. On the other hand, $\gamma(\boldsymbol{\mathrm{x}}) \to 0$ for $\boldsymbol{\mathrm{x}} \in \mathcal{B}$ implies $\ln \gamma(\boldsymbol{\mathrm{x}}) \to - \infty$ and $\mathcal{B}$ being a finite region, integral in the numerator of argument of Term 2 approaches negative of infinity which implies the term goes to 0. Thus, nFP goes to zero. The claim is only true for a finite region $\mathcal{B}$. This limiting case is used in section \ref{sec:density_eval} to show that the proposed density evaluation method (section \ref{sec:FPM}) allows for existence of pure 0-1 solutions in the design space.
	\subsection{Numerical evaluation of $G_n$}
	\label{sec:Num_eval}
	Section \ref{sec:PF} established an analytical formula for product of a field over a domain. This section aims at establishing an exact method for numerical evaluation of eqn. \ref{eqn:Field_prod}. Topology optimization usually solves for piece-wise constant approximation of the concerned fields, see section \ref{sec:formulation_def}. Therefore, we take a piece-wise constant $\gamma(\boldsymbol{\mathrm{x}})$ for numerical integration.

	Let the partition $P = \{\Omega_1, \Omega_2, \hdots \Omega_n\}$ of $\Omega$ where $n$ is finite, be such that $\gamma(\boldsymbol{\mathrm{x}}) = \gamma_i$, a constant, for $\boldsymbol{\mathrm{x}} \in \Omega_i$. Then, $G_n$ in eqn. \ref{eqn:Field_prod} can be simplified as,
	\begin{eqnarray}\nonumber
	G_m = \exp \left( \dfrac{\displaystyle\int\limits_{\Omega} \ln \gamma(\boldsymbol{\mathrm{x}}) d\Omega}{\displaystyle\int\limits_{\Omega} d\Omega} \right) &=& \exp \left( \dfrac{\displaystyle\sum\limits_{i = 1}^{n} \displaystyle\int\limits_{\Omega_i} \ln \gamma(\boldsymbol{\mathrm{x}}) d\Omega}{\displaystyle\int\limits_{\Omega} d\Omega} \right)\\
	&=& \exp \left(\displaystyle\sum\limits_{i = 1}^{n} \dfrac{ (\ln \gamma_i) A(\Omega_i)}{A(\Omega)} \right)
	\label{eqn:num_eval_1}\\
	&=& \prod_{i = 1}^{n} \gamma_i^{A(\Omega_i)/A(\Omega)}
	\label{eqn:num_eval_2}
	\end{eqnarray}
	where $A(\Omega_i)$ and $A(\Omega)$ are the area or volume of $\Omega_i$ and $\Omega$ respectively. Thus, the desired expression can be evaluated as an exponentiated product of the various values $\gamma(\boldsymbol{\mathrm{x}})$ takes over the domain.

	In what follows, we consider a special case where $A(\Omega_i) = A(\Omega_j)~ \forall i,j \in \{1,2, \hdots, n\}$ and $0 < \gamma_i < 1$. Issues in accurate evaluation of eqn. \ref{eqn:num_eval_2} arise when $\gamma_i$ is below  machine precision. For example, consider when $\Omega$ is a rectangular region divided into 25 partitions of equal areas, that is, $n = 25$ and $A(\Omega_i) = A(\Omega_j)$, and let the machine precision be $10^{-16}$. Let $\gamma_1 = 10^{-20}$, $\gamma_2 = 10^{-5}$ and $\gamma_i = 1$ for $i \in \{3,4, \hdots 24,25\}$. Then, the evaluation of MFP of $\gamma(\boldsymbol{\mathrm{x}})$ over $\Omega$ using eqn. \ref{eqn:num_eval_2},
	\begin{eqnarray}\nonumber
	G_n = \prod_{i = 1}^{25} \gamma_i^{1/25} = 0.1,
	\end{eqnarray}
	can potentially have numerical errors as $\gamma_1$ is below machine precision.
	Such scenarios often arise in the formulation discussed ahead. Thus, it is recommended to use $\ln \gamma_i$ instead of $\gamma_i$ and employ eqn. \ref{eqn:num_eval_1} instead of eqn. \ref{eqn:num_eval_2} for evaluation of nFP. An added advantage of implementing  $\ln\gamma_i$ is that it allows for $\gamma_i$ to take values much below machine precision such as $\gamma_i = 10^{-50}$. Another advantage of working with $\ln \gamma_i$ is discussed in section \ref{sec:sensitivity}.

%	\pagebreak
%	
	\section{Density evaluation methods}
	\label{sec:formulation_def}
	This section first lays out some criteria an ideal topology optimization
	formulation is expected to fullfil. Such criteria are mentioned previously by many practitioners, e.g., [Sig07]. This is followed by some definitions required to discuss density based topology optimization formulations. Some existing topology optimization formulations are then assessed under these criteria and a potential alternate approach is introduced. Note that symbols and definitions herein are independent of those used in section \ref{sec:Math}.\\
	Density methods aim at determining the density distribution, $0 \leq \rho(\boldsymbol{\mathrm{X}}) \leq 1$, over a design domain $\Omega$, that is, $\boldsymbol{\mathrm{X}} \in \Omega$ to minimize a prescribed objective. $\rho(\boldsymbol{\mathrm{X}}) = 1$ and $\rho(\boldsymbol{\mathrm{X}}) = 0$ imply presence and absence of material at $\boldsymbol{\mathrm{X}} \in \Omega$ respectively. An ideal topology optimization formulation should 
	\begin{enumerate}[a.]
		\item allow for pure 0-1 solutions, that is, $\rho(\boldsymbol{\mathrm{X}}) = 0~ \text{or}~ 1$ throughout the domain,
		\item ensure existence of solution, that is, yield mesh independent solutions and
		\item be free of user-defined parameters.
%		\item ensures that the evaluation of $\boldsymbol{\rho}$ from $\boldsymbol{\alpha}$ is independent of the domain discretization.
	\end{enumerate}
	The above list is a sub-set of qualities of an ideal topology optimization formulation presented in \cite{sigmund2007morphology}. Many practitioners have proposed methods to achieve most of these objectives for small deformation continua. We present three well practiced and one potential topology optimization formulation, and discuss them under the aforementioned criteria. We first establish some definitions on neighborhood and auxiliary field.

\begin{figure}[!htb]
		\begin{minipage}{0.6\textwidth}
			\centering
			\includegraphics[width=0.8\textwidth]{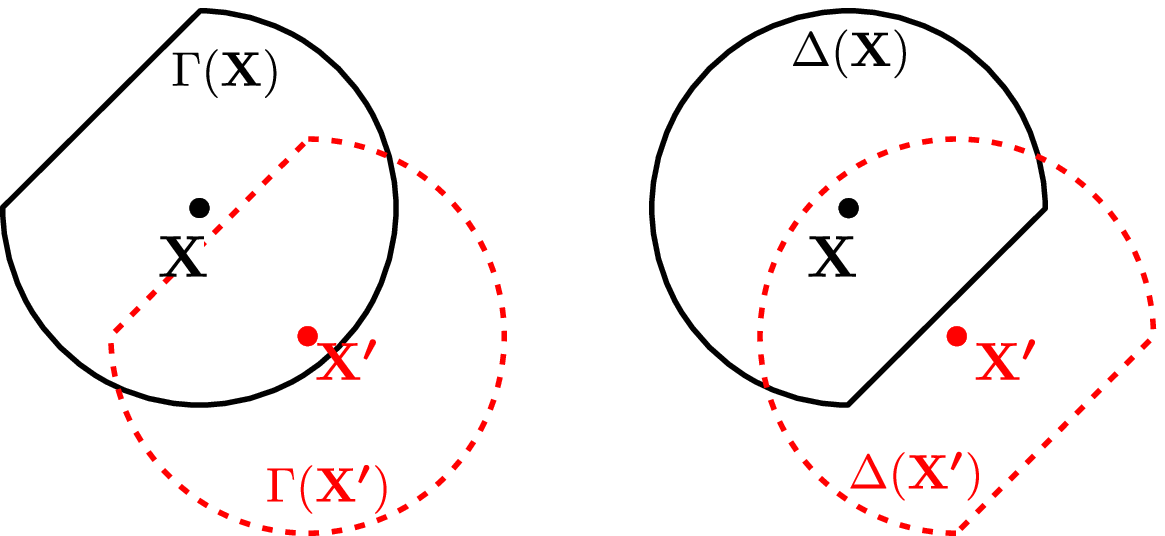}
			\caption{Neighborhood, $\Gamma(\boldsymbol{\mathrm{X}})$, and the corresponding $\Delta(\boldsymbol{\mathrm{X}})$. }
			\label{fig:neighborhood}
		\end{minipage}\hfill
		\begin{minipage}{0.45\textwidth}
			\centering
			\includegraphics[width=0.6\textwidth]{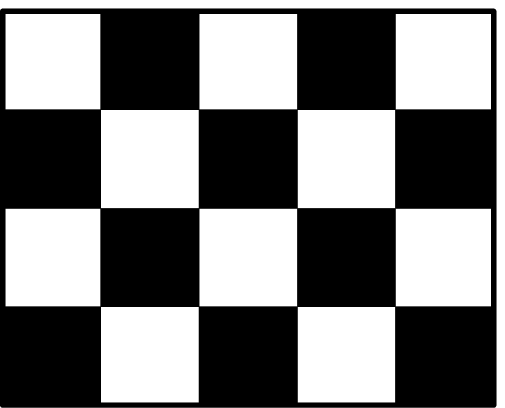}
			\caption{Checkerboard pattern }
			\label{fig:checkerboard}
		\end{minipage}
	\end{figure}

	Let $\Gamma(\boldsymbol{\mathrm{X}})$, referred to as the neighborhood of $\boldsymbol{\mathrm{X}}$, be a finite region surrounding $\boldsymbol{\mathrm{X}}$. Let $\Delta(\boldsymbol{\mathrm{X}})$ be the region enclosing all points for which $\boldsymbol{\mathrm{X}}$ lies in the neighborhood. For discussions here, the shape of $\Gamma(\boldsymbol{\mathrm{X}})$ is kept same for all $\boldsymbol{\mathrm{X}}$ thus making $\Delta(\boldsymbol{\mathrm{X}})$ a complimentary shape to $\Gamma(\boldsymbol{\mathrm{X}})$. An example of $\Gamma(\boldsymbol{\mathrm{X}})$ and its corresponding $\Delta(\boldsymbol{\mathrm{X}})$ is presented in fig. \ref{fig:neighborhood} for two arbitrary points $\boldsymbol{\mathrm{X}}$ and $\boldsymbol{\mathrm{X}'}$. In fig. \ref{fig:neighborhood}, $\boldsymbol{\mathrm{X}'} \in \Gamma(\boldsymbol{\mathrm{X}})$ but $\boldsymbol{\mathrm{X}} \notin \Gamma(\boldsymbol{\mathrm{X}}')$ and hence, $\boldsymbol{\mathrm{X}} \in \Delta(\boldsymbol{\mathrm{X}}')$ while $\boldsymbol{\mathrm{X}}' \notin \Delta(\boldsymbol{\mathrm{X}})$. Note that $\Delta(\boldsymbol{\mathrm{X}}) = \Gamma(\boldsymbol{\mathrm{X}})$ for $\Gamma(\boldsymbol{\mathrm{X}})$ point symmetric\footnote{A shape being point symmetric about a point $\boldsymbol{\mathrm{X}}$ implies that for any $\boldsymbol{\mathrm{r}}$, if $\boldsymbol{\mathrm{X}} + \boldsymbol{\mathrm{r}}$ lies on the boundary then $\boldsymbol{\mathrm{X}} - \boldsymbol{\mathrm{r}}$ also lies on the boundary.} about $\boldsymbol{\mathrm{X}}$.

	Most density based topology optimization formulations introduce an auxiliary field, $\alpha(\boldsymbol{\mathrm{X}}), \boldsymbol{\mathrm{X}} \in \Omega$, which by themselves do not have a physical interpretation. The intent is to express $\rho(\boldsymbol{\mathrm{X}})$ in terms of $\alpha(\boldsymbol{\mathrm{X}})$ and solve for $\alpha(\boldsymbol{\mathrm{X}})$, making $\alpha(\boldsymbol{\mathrm{X}})$ the primary variable. Formulations vary in their choice of expression connecting $\rho(\boldsymbol{\mathrm{X}})$ and $\alpha(\boldsymbol{\mathrm{X}})$. The choice of expression allows for implicit imposition of restrictions on the nature of $\rho(\boldsymbol{\mathrm{X}})$, as will be discussed ahead.

	All solution methods construct a numerical framework and attempt to achieve an approximation of the density field in a discretized version of the problem. Therefore, we transfer the above definitions and problem statement to a numerical setup. Let the domain $\Omega$ be arbitrarily divided into $n$ finite sized regions called \textit{cells}. The region enclosed within the $i^{th}$ cell is represented by $\Omega_i$ and its centroid by $\boldsymbol{\mathrm{X}}_i$. The fields $\rho(\boldsymbol{\mathrm{X}})$ and $\alpha(\boldsymbol{\mathrm{X}})$ are represented by their piece-wise constant approximation, constant within each cell, thus expressed using the $n$-dimensional vectors $\boldsymbol{\rho} = \{\rho_1~ \rho_2 \hdots \rho_n\}$ and $\boldsymbol{\alpha} = \{\alpha_1~ \alpha_2 \hdots \alpha_n\}$ respectively, where $\rho_i \equiv \rho(\boldsymbol{\mathrm{X}}_i)$ and $\alpha_i \equiv \alpha(\boldsymbol{\mathrm{X}}_i)$. Let $\{\mathbb{N}_i\}$ and $\{\mathbb{D}_i\}$ be the set of cells contained within $\Gamma_i \equiv \Gamma(\boldsymbol{\mathrm{X}}_i)$ and $\Delta_i \equiv \Delta(\boldsymbol{\mathrm{X}}_i)$ respectively. That is, $\{\mathbb{N}_i\} = \{~j~ |~ \boldsymbol{\mathrm{X}}_j \in \Gamma_i\}$ and $\{\mathbb{D}_i\} = \{~j~ |~ \boldsymbol{\mathrm{X}}_j \in \Delta_i\}$.

	To propose an expression connecting $\boldsymbol{\rho}$ to $\boldsymbol{\alpha}$ in the numerical framework, it is imperative to ensure that the evaluation of $\boldsymbol{\rho}$ from $\boldsymbol{\alpha}$ is independent of the domain discretization.
	This independence implies that for a prescribed field $\alpha(\boldsymbol{\mathrm{X}})$, evaluation of $\rho(\boldsymbol{\mathrm{X}})$ is independent of domain discretization and thus the two fields can possibly be related analytically. This idea braces no association with the concept of mesh independent solutions in topology optimization which states that the optimal solution for $\rho(\boldsymbol{\mathrm{X}})$ should be independent of the underlying discretization. We present and discuss four possible expressions for connecting $\boldsymbol{\rho}$ to $\boldsymbol{\alpha}$ and analyze them under the aforementioned criteria. In all cases, we take $0 \leq \alpha_i \leq 1$.
	\begin{description}
		\item[Case 1:] We allow for element densities ($\rho_i$) to take values independent of each other. That is, there is no implicit or explicit connection between any two element densities. This is achieved by expressing
		\begin{eqnarray}
		\rho_i = \alpha_i
		\label{eqn:density_method}
		\end{eqnarray} 
		where $\alpha_i$ are independent variables. The above expression allows for pure 0-1 solutions and leads to a discretization independent evaluation of element densities, connecting the two fields analytically as $\rho(\boldsymbol{\mathrm{X}}) = \alpha(\boldsymbol{\mathrm{X}})$. This formulation is extensively studied in literature and is shown to give mesh dependent results implying a solution may not exist. Loosely stating, existence of solution can be ensured by restricting rapid variations in the density field. Here, rapid variation refers to element densities alternating between 0 and 1 among adjacent cells. A well known example is the checker board pattern (Fig. \ref{fig:checkerboard}). These patterns often present themselves in compliance minimization and related problems when solved using rectangular meshes. Eqn. \ref{eqn:density_method} only allows for checker board patterns and does not enforce it. Their appearance in compliance minimization problems is a consequence of numerical and connectivity anomalies \citep{jog1996stability, diaz1995checkerboard}. One of the widely used approaches to arrest rapid density variation is discussed below.
		\item[Case 2:] The possibility of rapid variation in element densities is removed by introducing a sense of smoothness, achieved by expressing $\rho_i$ as \citep{bruns2001topology},
		\begin{eqnarray}
		\rho_i = \dfrac{\displaystyle\sum\limits_{j \in \{\mathbb{N}_i\}} w_j\alpha_j A(\Omega_j)}{\displaystyle\sum\limits_{j \in \{\mathbb{N}_i\}} w_j A(\Omega_j)} 
		\label{eqn:num_filter}
		\end{eqnarray} 
		where $A(\Omega_j)$ is the area or volume of cell $\Omega_j$ and $w_j > 0$ are user defined weights. Thus element densities are a weighted sum of the discrete auxiliary field in their neighborhood. An implicit control in density variation among adjacent cells is achieved by defining neighborhoods such that the latter have considerable overlap. For the case described above, the two fields, in the continuum setting, are linked by the analytical relation
		\begin{eqnarray}
			\rho(\boldsymbol{\mathrm{X}}) = \dfrac{\displaystyle\int\limits_{\Gamma(\boldsymbol{\mathrm{X}})} w(\boldsymbol{\mathrm{X}})\alpha(\boldsymbol{\mathrm{X}}) dA}{\displaystyle\int\limits_{\Gamma(\boldsymbol{\mathrm{X}})}  w(\boldsymbol{\mathrm{X}})dA}.
			\label{eqn:analy_filter}
		\end{eqnarray}
		The above description is a generic case of the filtering process \citep{bourdin2001filters} where the auxiliary fields, $\alpha(\boldsymbol{\mathrm{X}})$, is commonly referred to as the unfiltered density field, and $w(\boldsymbol{\mathrm{X}})$ is a user defined weight function inversely proportional to the distance of a point from $\boldsymbol{\mathrm{X}}$. Traditionally, in filtering, $\Gamma(\boldsymbol{\mathrm{X}})$ is taken as a circular region centered at $\boldsymbol{\mathrm{X}}$ with a pre-specified radius. Filtering formulations vary in their choice of the weight function. Filtering is widely used in topology optimization \citep{bruns2003element, wang2005bilateral} and is known to yield mesh independent solutions. \cite{bourdin2001filters} proves existence of solution for the topology optimization problem for the general case of filtering. Drawback of the filtering process is that a transition from solid to void phase inevitably goes through a region of intermediate density. Depending on the weight function and the filtering radius, the phase transition may occur over multiple cells, thus removing the possibility of pure 0-1 solutions.
		\item[Case 3:] Another approach to avoid rapid variations in density is to introduce a sense of length scale in the formulation. Filtering also adopts this idea as the density of a cell $i$, $\rho_i$, is influenced by $\alpha_j$, $j \in \{\mathbb{N}_i\}$, but the contribution/influence of an $\alpha_j$ on $\rho_i$ is fairly limited. %For the choice of weigh functions usually implemented, the maximum contribution of $\alpha_j$ to $\rho_i$ reduces as one moves away from the centroid.
		To overcome this shortcoming, \cite{Guest2004} introduces the projection method which associates the auxiliary field, $\alpha(\boldsymbol{\mathrm{X}})$, with a material phase (solid or void). An active auxiliary field $(\alpha(\boldsymbol{\mathrm{X}}) > 0)$ projects the associated material phase onto all elements in the neighborhood using a heavyside function. For solid phase projection, the density field is expressed as
		\begin{eqnarray}
		\rho_i = 1 - e^{-\beta \mu_i} + \mu_i e^{-\beta} 
		\label{eqn:num_projection}
		\end{eqnarray}
		where $\beta > 0$ is a user defined parameter and  
		\begin{eqnarray}\nonumber
		\mu_i = \dfrac{\displaystyle\sum\limits_{j \in \{\mathbb{D}_i\}} w_j\alpha_j A(\Omega_j)}{\displaystyle\sum\limits_{j \in \{\mathbb{D}_i\}} w_j A(\Omega_j)}.
		\end{eqnarray}
		is similar to the filtered density in eqn. \ref{eqn:num_filter}.
		The expression above, however, takes summation over $\{\mathbb{D}_i\}$ unlike eqn. \ref{eqn:num_filter} which takes summation over $\{\mathbb{N}_i\}$ (see case 2). As projection usually implements a circular $\Gamma(\boldsymbol{\mathrm{X}})$, $\{\mathbb{D}_i\} = \{\mathbb{N}_i\}$. Projection is shown to provide mesh independent solutions and connect the two fields, in the continuum setting, by the analytical formula
		\begin{flalign}\label{eqn:analy_projection}
		&~&\rho(\boldsymbol{\mathrm{X}}) &= 1 - e^{-\beta \mu(\boldsymbol{\mathrm{X}})} + \mu(\boldsymbol{\mathrm{X}}) e^{-\beta}&\\[8pt]\nonumber
		&\text{where}& 
		\mu(\boldsymbol{\mathrm{X}}) &=  \dfrac{\displaystyle\int\limits_{\Delta(\boldsymbol{\mathrm{X}})} w(\boldsymbol{\mathrm{X}})\alpha(\boldsymbol{\mathrm{X}}) dA}{\displaystyle\int\limits_{\Delta(\boldsymbol{\mathrm{X}})}  w(\boldsymbol{\mathrm{X}})dA}.&
%		\label{eqn:analy_mu}
		\end{flalign}
		As one increases $\beta$, the transition region can be made narrow. Thus, projection can provide close to binary solutions for a high value of $\beta$ but a pure 0-1 solution is still not part of its solution space. Also, as the mesh is refined, a higher value of $\beta$ is required to close in on 0-1 solutions. As, high values of $\beta$ adversely affect the gradient magnitudes, proper implementation of the projection method requires a heuristic based continuation approach on parameter $\beta$ to obtain desirable solutions.
		
		\item[Case 4:] With the intent to incorporate pure 0-1 solutions within the solution space, for the novel fourth case, we develop an expression such that if $\alpha_i = 1$, then $\rho_j = 1~ \forall~ j \in \{\mathbb{N}_i\}$, while $\rho_i = 0$ implies that $\alpha_j = 0 ~ \forall j \in \{\mathbb{D}_i\}$. We assign neighborhood $\Gamma_i = \Gamma(\boldsymbol{\mathrm{X}}_i)$ at each $\boldsymbol{\mathrm{X}}_i$. Depending on value the auxiliary field $\alpha(\boldsymbol{\mathrm{X}})$ takes at $\boldsymbol{\mathrm{X}}_i$, each $\Gamma_i$ can be dormant ($\alpha_i = 0$), partially active ($0 <\alpha_i < 1$) or active ($\alpha_i = 1$). An active neighborhood, $\Gamma_i$ ensures that the density of all cells enclosed within it is 1 while a dormant neighborhood has no impact on the densities. Thus, $\alpha_i$ impacts the density for all cells in $\{\mathbb{N}_i\}$ while the density at $\boldsymbol{\mathrm{X}}_i$, $\rho_i$, is impacted by the auxiliary field at all cells for which $\boldsymbol{\mathrm{X}}_i$ lies in the neighborhood, $\{\mathbb{D}_i\}$. This can be achieved by expressing $\rho_i$ as,
		\begin{eqnarray}
		\rho_i = 1 - \prod_{j \in \{\mathbb{D}_i\}} (1 - \alpha_j).
		\label{eqn:BPP}
		\end{eqnarray}
		The above relation allows for pure 0-1 topological solutions in $\rho_i$ irrespective of the mesh size, a feature that existing filtering (case 2) or projection (case 3) methods do not allow for. Fig. \ref{fig:density_dis_schem}b presents the density distribution for an example auxiliary field in fig. \ref{fig:density_dis_schem}a over a domain discretized using rectangular cells for the case when $\{\mathbb{N}_i\}$ contains only the immediate neighbors of a cell. Fig. \ref{fig:density_dis_schem}a depicts neighborhood of a point $\boldsymbol{\mathrm{P}}$ using hashed lines with negative slope while $\Delta$ for points $\boldsymbol{\mathrm{Q}}$ and $\boldsymbol{\mathrm{R}}$ are highlighted using hashed lines with positive slope. Cells with $\alpha = 1$ are highlighted in gray in fig. \ref{fig:density_dis_schem}a and cells with $\rho = 1$ are highlighted in black in fig. \ref{fig:density_dis_schem}b. As demonstrated, the above expression can lead to single cell voids (regions with $\rho = 0$) but not single cell solids (regions with $\rho = 1$). Unlike filtering and projection methods, the density formulation in eqn. \ref{eqn:BPP} is independent of any function or parameter choices. The issue of mesh independent solutions remains unexplored for the above expression.

		It is evident that eqn. \ref{eqn:BPP} does not lead to discretization independent evaluation of element densities. To show this consider the case of constant auxiliary field $\alpha(\boldsymbol{\mathrm{X}}) = \alpha_0$ for a given mesh refinement, $\mathrm{R}$. Let $r$ be the size of $\{\mathbb{D}_i\}$ for $\mathrm{R}$. Then, $\rho_i$ from eqn. \ref{eqn:BPP} would be $\rho_i = 1 - (1- \alpha_0)^r$. As the evaluation is dependent on $r$ which varies with mesh refinement, $\rho_i$ from eqn. \ref{eqn:BPP} gives discretization dependent results.
	\end{description}
	\begin{figure}[!htb]
		\centering
		\subcaptionbox{Cells with $\alpha_i = 1$ are highlighted in gray. Neighborhood of point $\boldsymbol{\mathrm{P}}$ is hashed.}{\includegraphics[width=0.4\textwidth]{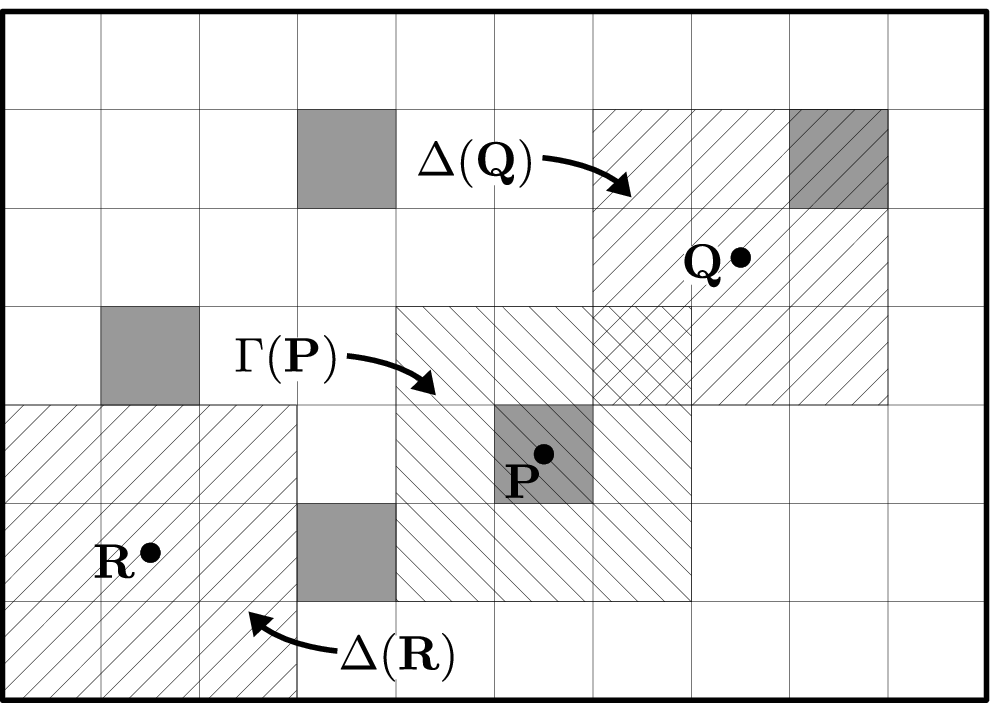}}
		\label{fig:density_dis_schem_a}%
		\hspace{0.7cm}
		\subcaptionbox{Density distribution for the auxiliary field Cells with $\rho_i = 1$ are highlighted in black.}{\includegraphics[width=0.4\textwidth]{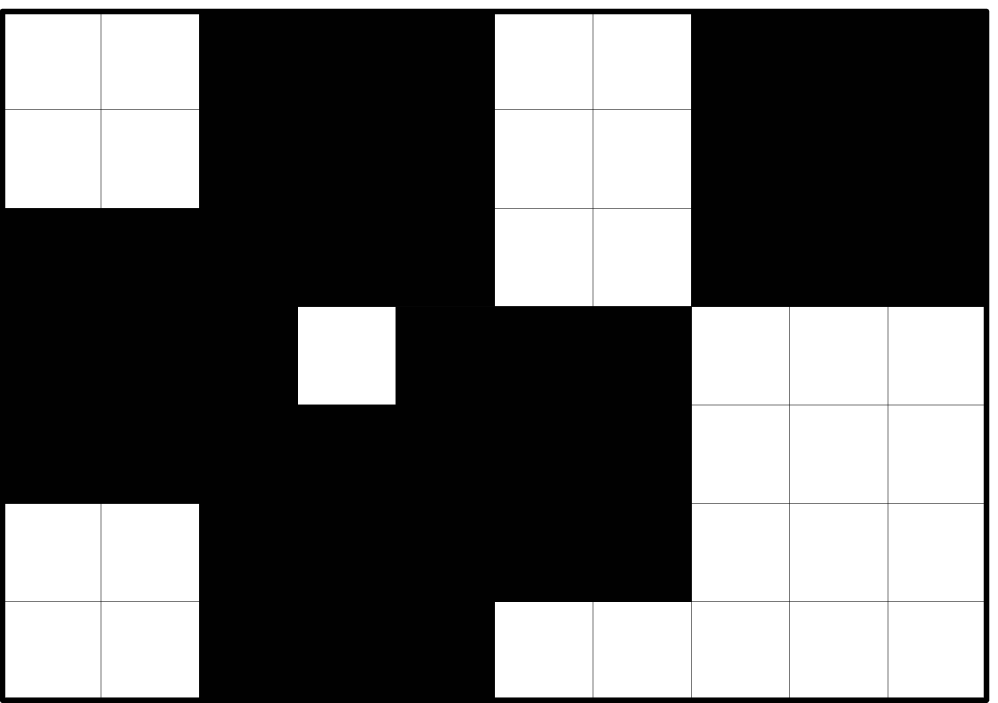}}
		\label{fig:density_dis_schem_b}%
		\caption{Density distribution schematic for case 4.}
		\label{fig:density_dis_schem}
	\end{figure}
	It is expected that the expression in eqn. \ref{eqn:BPP} can lead to mesh independent and binary solutions as it removes the possibility of rapid density variations and also, binary solutions get included within the design space. However, the expression leads a mathematically incoherent optimization problem as the density evaluation from auxiliary field depends on mesh refinement. Thus, an attempt to modify eqn. \ref{eqn:BPP} such that an analytical connection between $\rho(\boldsymbol{\mathrm{X}})$ and $\alpha(\boldsymbol{\mathrm{X}})$ can be established is made. From observation, similar to eqn. \ref{eqn:num_filter} which presents element density as weighted sum of a field, eqn. \ref{eqn:BPP} attempts to expresses element density as product of a field over a region. Building on this observation, the analytical formula for product of a field, presented earlier (section \ref{sec:Math}), is developed.

%	\pagebreak
%	
	\section{Normalized Field Product method for topology optimization}
\label{sec:FPM}
	In this section, the analytical expression for product of a field (eqn. \ref{eqn:num_eval_2}) is utilized to modify eqn. \ref{eqn:BPP} to develop a novel density formulation in topology optimization to include pure 0-1 solutions within the design space. The formulation thus proposed is discussed under the criteria for an ideal density evaluation method, presented in section \ref{sec:formulation_def}. Further, sensitivity analysis for a generic objective is presented followed by a discussion on the same. %Note that the definitions and symbols implemented in this section are same as that of section \ref{sec:formulation_def} and independent of section \ref{sec:Math}.
	\subsection{Density evaluation}
	\label{sec:density_eval}
	Taking inspiration from eqn. \ref{eqn:num_eval_2}, eqn. \ref{eqn:BPP} is modified in that, the cell density $\rho_i$ is expressed as an nFP of $\alpha (\boldsymbol{\mathrm{X}})$ over $\Delta_i$ as follows,
	\begin{eqnarray}
	\rho(\boldsymbol{\mathrm{X}}_i) \equiv \rho_i = 1 - \prod_{j \in \{\mathbb{D}_i\}} (1 - \alpha_j)^{A(\Omega_j)/A(\Delta_i)}
	\label{eqn:BPP_mod}
	\end{eqnarray}
	where $A(\Delta_i)$ is the area or volume of the region $\Delta_i$ evaluated as,
	\begin{eqnarray}\nonumber
	A(\Delta_i) = \sum\limits_{j \in \{\mathbb{D}_i\}} A(\Omega_j).
	\end{eqnarray}
	Eqn. \ref{eqn:BPP_mod} introduces an exponent in eqn. \ref{eqn:BPP}, doing which eliminates the feature of discretization dependent evaluation of $\boldsymbol{\rho}$ from $\boldsymbol{\alpha}$ connecting the two fields, $\alpha(\boldsymbol{\mathrm{X}})$ and $\rho(\boldsymbol{\mathrm{X}})$, via the analytical expression 
	\begin{eqnarray}
	\rho(\boldsymbol{\mathrm{X}}) = 1 - \exp \left( \dfrac{\displaystyle\int\limits_{\Delta(\boldsymbol{\mathrm{X}})} \ln (1- \alpha(\boldsymbol{\mathrm{X}})) dA}{\displaystyle\int\limits_{\Delta(\boldsymbol{\mathrm{X}})} dA} \right).
	\label{eqn:density_formula}
	\end{eqnarray}
	Eqn. \ref{eqn:BPP_mod} is now analyzed under the same criteria as were the density expressions in eqns. \ref{eqn:density_method}, \ref{eqn:num_filter} and \ref{eqn:num_projection}.\\
	As the two fields, $\alpha(\boldsymbol{\mathrm{X}})$ and $\rho(\boldsymbol{\mathrm{X}})$, can be connected in a continuum setting, evaluation of cell densities is independent of the discretization. \\
	It is evident from eqn. \ref{eqn:BPP_mod} that $\alpha_j=1$ for any $j \in \{\mathbb{D}_i\}$ gives $\rho_i = 1$ but the corresponding continuum relation, eqn. \ref{eqn:density_formula}, does not allow for $\alpha(\boldsymbol{\mathrm{X}}) = 1$. Thus it is essential to look at the limiting case of $\alpha_j \to 1$. As $\boldsymbol{\alpha}$ is a piece-wise constant approximation of $\alpha(\boldsymbol{\mathrm{X}})$, piece-wise constant within each cell, $\alpha_j \to 1$ implies $\alpha(\boldsymbol{\mathrm{X}}) \to 1$ for $\boldsymbol{\mathrm{X}} \in \Omega_j$.
	Taking $\gamma(\boldsymbol{\mathrm{X}}) = 1 - \alpha(\boldsymbol{\mathrm{X}})$, $\mathcal{B} = \Omega_j$, $j \in \{\mathbb{D}_i\}$, and replacing $\Omega$ by $\Delta(\boldsymbol{\mathrm{X}})$ in eqn. \ref{eqn:conv_to_zero}, we observe that if $\alpha(\boldsymbol{\mathrm{X}}) \to 1$ within any cell inside $\Delta(\boldsymbol{\mathrm{X}})$, then the density at $\boldsymbol{\mathrm{X}}$, $\rho(\boldsymbol{\mathrm{X}})$, also approaches 1. This implies that in eqn. \ref{eqn:BPP_mod} if $\alpha_j \to 1$ for any $j \in \{\mathbb{D}_i\}$ then $\rho_i \to 1$. Note that, unlike eqn. \ref{eqn:BPP_mod} where $\rho_i = 1$ is possible, that is, binary solutions are within the design space, eqn. \ref{eqn:density_formula} does not allow for $\rho(\boldsymbol{\mathrm{X}}) = 1$. Thus eliminating pure 0-1 solutions from the solution space. However, $\rho(\boldsymbol{\mathrm{X}})$ can be brought sufficiently close to 1 for practical purposes. On the other hand, $\rho(\boldsymbol{\mathrm{X}}) = 0$ is achieved when, $\alpha(\boldsymbol{\mathrm{X}}) = 0~ \forall ~\boldsymbol{\mathrm{X}} \in \Delta(\boldsymbol{\mathrm{X}})$, that is, for $\rho_i = 0$, $\alpha_j = 0 ~\forall ~j \in \{\mathbb{D}_i\}$ must hold. These properties match exactly with those of eqn. \ref{eqn:BPP}. It is therefore deduced that similar to eqn. \ref{eqn:BPP}, eqn. \ref{eqn:BPP_mod} allows for black and white solutions. However it, does not enforce it. That is, there is no guarantee of getting a pure solid-void solution.\\
	Hence, the density evaluation method in eqn. \ref{eqn:BPP_mod} allows for 0-1 solutions, is free of any function or parameter choices and gives discretization independent evaluation of densities, thus satisfying two of the three criteria mentioned in section \ref{sec:formulation_def}. Note that, implementing eqn. \ref{eqn:BPP_mod} requires the definition for $\Delta(\boldsymbol{\mathrm{X}})$ which in turn requires a choice for the neighborhood, $\Gamma(\boldsymbol{\mathrm{X}})$, to be made. Similar to filtering and projection method, $\Gamma(\boldsymbol{\mathrm{X}})$ imposes an implicit length scale requirement on the density distribution. Satisfaction of the final criteria, mesh independent solutions, requires a rigorous proof for existence of solution which is beyond the scope of this work. However, to demonstrate mesh independence, solutions to a few standard topology optimization problems for varying mesh sizes using the aforementioned formulation are presented in section \ref{sec:results}. \\
	Before expressing the topology optimization formulation for field product method, a final modification to eqn. \ref{eqn:density_formula} is made. We set $\beta(\boldsymbol{\mathrm{X}}) = \ln(1- \alpha(\boldsymbol{\mathrm{X}}))$ as it allows for $1-\alpha(\boldsymbol{\mathrm{X}})$ to take values much lower than machine precision (sec. \ref{sec:Num_eval}). Eqn. \ref{eqn:density_formula} and eqn. \ref{eqn:BPP_mod} are modified accordingly as
	\begin{flalign}\label{eqn:FPM_analy}
	&~& \rho(\boldsymbol{\mathrm{X}}) &= 1 - \exp \left( \dfrac{\displaystyle\int\limits_{\Delta(\boldsymbol{\mathrm{X}})} \beta(\boldsymbol{\mathrm{X}}) dA}{\displaystyle\int\limits_{\Delta(\boldsymbol{\mathrm{X}})} dA} \right)& \\[8pt]	
	&\text{and}& \rho_i &= 1 - \exp \left(\dfrac{\sum\limits_{j \in \{\mathbb{D}_i\}}\beta_j A(\Omega_j)}{A(\Delta_i)}\right)&
	\label{eqn:FPM_num}
	\end{flalign}
	respectively, where $-\infty < \beta(\boldsymbol{\mathrm{X}}) \leq 0$. Similar to as eqn. \ref{eqn:density_formula}, eqn. \ref{eqn:FPM_num} does not allow for $\rho_i = 1$, but large values of $\beta_j$ can approach $1$ close enough to machine precision. Thus, by pure 0-1 solution we imply that $\rho_i = 0$ or $\rho_i \approx 1~ \forall~ i \in \{1,2,3,....,ne\}$. Note that a topology with $\rho(\boldsymbol{\mathrm{X}}) = 0$ or $ \rho(\boldsymbol{\mathrm{X}}) \approx 1 ~\forall ~ \boldsymbol{\mathrm{X}} \in \Omega$ is not possible. Eqn. \ref{eqn:FPM_analy} enforces that $\rho(\boldsymbol{\mathrm{X}})$ is continuous even for a piece-wise continuous $\beta(\boldsymbol{\mathrm{X}})$. Thus for the analytical field $\rho(\boldsymbol{\mathrm{X}})$, there is no jump possible and thus, there will always exist a transition region at structural boundaries. A corollary of the discussions in section \ref{sec:conv_zero} is that these transition regions can be infinitesimally small. This is not possible in projection or filtering by design as weighted average implemented in those methods will always take a finite transition region.

	Appendix \ref{Appendix:A} addresses bijectivity between the two scalar fields and discusses a way to evaluate the design variables, $\boldsymbol{\beta}$, for a given density distribution, $\boldsymbol{\rho}$. The discussion pertains to only academic completeness and is not a requirement for implementation of the proposed nFP method.

	\subsection{Optimization formulation}
	Various features of the normalized field product method for topology optimization are showcased via solutions to a generic small deformation continuum optimization problem.
	\begin{flalign}\label{eqn:2}
	&~& \text{minimize} &~~~~ \mathbb{I}(\boldsymbol{\rho}) &\\ \nonumber
	&\text{subject to} & \bm{\mathrm{R}}(\boldsymbol{\rho},\bm{\mathrm{u}})  &= \bm{0} &\\ \nonumber
	&\text{such that}& \sum_{i=1}^{ne} \rho_i(\boldsymbol{\beta}) v_i &\leq V^* = v_f \sum_{i=1}^{ne} v_i & \\ \nonumber
	& ~ & \beta_{lb} \leq \beta_i & \leq 0 &
	\end{flalign}
	where $\mathbb{I}(\boldsymbol{\rho})$ is the objective, $\bm{\mathrm{R}}(\boldsymbol{\rho},\bm{\mathrm{u}})$ is the state equation to be satisfied for any intermediate continuum, $\boldsymbol{\beta} = \{\beta_1 ~\beta_2 \hdots \beta_{ne}\}$ is the piece-wise constant approximation of $\beta(\boldsymbol{\mathrm{X}})$, $v_i$ is the geometric volume of element $i$, $v_f$ is the user specified volume fraction, $ne$ is the total number of elements in the FE discretization of the domain and $\beta_{lb} < 0$ is the lower bound on $\beta_i$. It is evident from eqn. \ref{eqn:FPM_num} that a lower bound on $\beta_i$ is not required for density evaluation, but introducing a lower bound on the design variables leads to a closed design space, ensuring existence of solution in the numerical framework of the problem (eqn. \ref{eqn:2}). The description above uses the same domain discretization for the FE analysis and approximation of the fields $\alpha(\boldsymbol{\mathrm{X}})$ and $\beta(\boldsymbol{\mathrm{X}})$. $\bm{\mathrm{R}}(\boldsymbol{\rho},\bm{\mathrm{u}}) = \bm{0}$ is the force balance equation $\bm{\mathrm{K}}(\boldsymbol{\rho}) \bm{\mathrm{u}} = \bm{\mathrm{F}}$, where $\bm{\mathrm{K}}(\boldsymbol{\rho})$ is the global stiffness matrix corresponding to $\boldsymbol{\rho}$, $\bm{\mathrm{u}}$ is the nodal displacement vector and $\boldsymbol{\mathrm{F}}$ is the external force vector. The objective function and stiffness matrix are functions of elemental densities which in turn can be expressed in terms of design variables $\boldsymbol{\beta}$ by  eqn. \ref{eqn:FPM_num}. The objective depends on the problem while, the relation between stiffness matrix and density depends on the choice of material model. We implement the SIMP material model \citep{bendsoe1999material}
	for which the elemental stiffness, $\boldsymbol{\mathrm{Ke}}$, for an element with density $\rho_i$ is given as,
	\begin{eqnarray}
	\label{eqn:SIMP}
	\boldsymbol{\mathrm{Ke}} = \{\rho_i^\eta (1-\rho_{min}) + \rho_{min}\}\boldsymbol{\mathrm{K}}_0
	\end{eqnarray}
	where $\eta$ is the SIMP penalty parameter, $\rho_{min}$ is a small positive number introduced to remove potential singularity of the stiffness matrix and $\boldsymbol{\mathrm{K}}_0$ is the elemental stiffness of a solid cell. 
	\subsection{Sensitivity analysis}
	\label{sec:sensitivity}
	Sensitivity analysis concerns itself with evaluating gradients of the objective, $\mathbb{I}(\boldsymbol{\rho})$, with respect to the design variable, $\boldsymbol{\beta}$. Implementing chain rule, we get
	\begin{eqnarray}
	\dfrac{\partial\mathbb{I}(\boldsymbol{\rho})}{\partial \beta_j} = 
	\sum\limits_{i=1}^{ne}~ \dfrac{\partial\mathbb{I}(\boldsymbol{\rho})}{\partial \rho_i} \times \dfrac{\partial\rho_i}{\partial \beta_j}
	\end{eqnarray}  
	where derivative of the objective with respect to element densities depends on the problem while derivative of element densities with respect to design variables $\beta_j$ is given as,
	\begin{eqnarray}
	\label{eqn:density_grad}
	\dfrac{\partial\rho_i}{\partial \beta_j} = 
	\begin{cases}
	-\dfrac{(1-\rho_i)A(\Omega_j)}{A(\Delta_i)} &\text{for}~ i \in \{\mathbb{N}_j\}\\
	0 &\text{for}~ i \notin \{\mathbb{N}_j\}.
	\end{cases}
	\label{eqn:gradient}
	\end{eqnarray}
	Gradients $\frac{\partial\rho_i}{\partial \beta_j}$ vary linearly with respect of element densities and are independent of the value $\beta_j$ takes. Also, $\frac{\partial \rho_i}{\partial \beta_j}$ is 0 for $\rho_i = 1$. Eqn. \ref{eqn:gradient} provides sufficient information to evaluate volume constraint gradient defined in eqn. \ref{eqn:2}.\\
	If one takes $\boldsymbol{\alpha}$ (eqn. \ref{eqn:density_formula}) as the design variables instead of $\boldsymbol{\beta}$, the gradients thus evaluated are singular at $\alpha_i=1$ while no such singularity appears in eqn. \ref{eqn:gradient}. This is an additional benefit of replacing $\boldsymbol{\alpha}$ by $\boldsymbol{\beta}$ as design variables in eqn. \ref{eqn:2}.

	\subsection{Domain discretization and Neighborhood}
	\label{sec:Boundary_consideration}
\begin{figure}[h]
		\centering
		\subcaptionbox{Discretization of $\Gamma_i$ for $ls = 1$}{\includegraphics[width=0.2\textwidth]{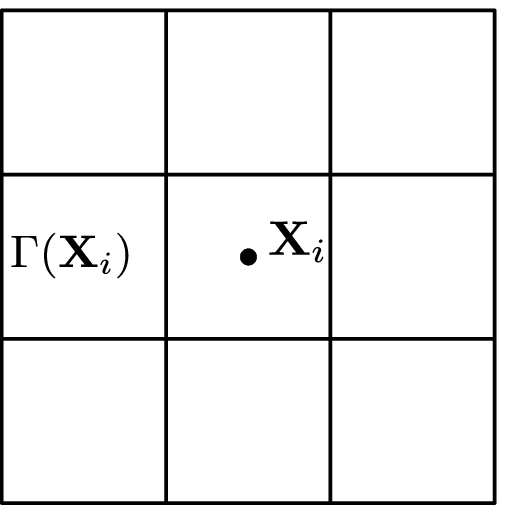}}
		\label{fig:Neighborhood_1}%
		\hspace{0.7cm}
		\subcaptionbox{Discretization of $\Gamma_i$ for $ls = 2$}{\includegraphics[width=0.2\textwidth]{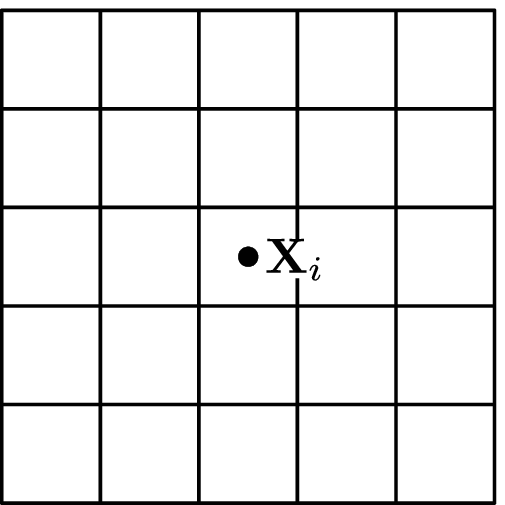}}
		\label{fig:Neighborhood_2}%
		\hspace{0.7cm}
		\subcaptionbox{Discretization of $\Gamma_i$ for $ls = 4$}{\includegraphics[width=0.2\textwidth]{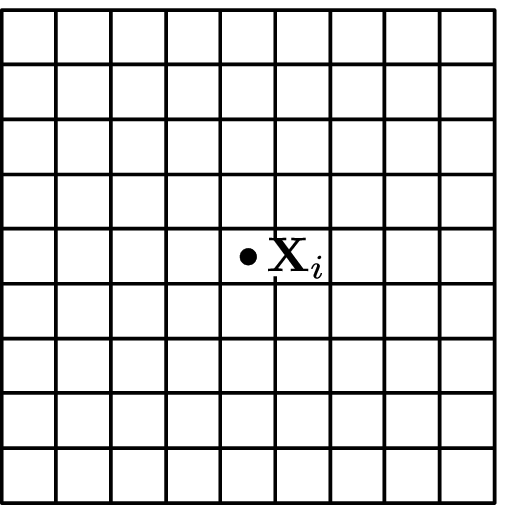}}
		\label{fig:Neighborhood_3}%
		%	\hspace{0.35cm} 
		\\[12pt]
		\subcaptionbox{Schematic of the neighborhood for interior and close to  boundary points.}{\includegraphics[width=0.4\textwidth]{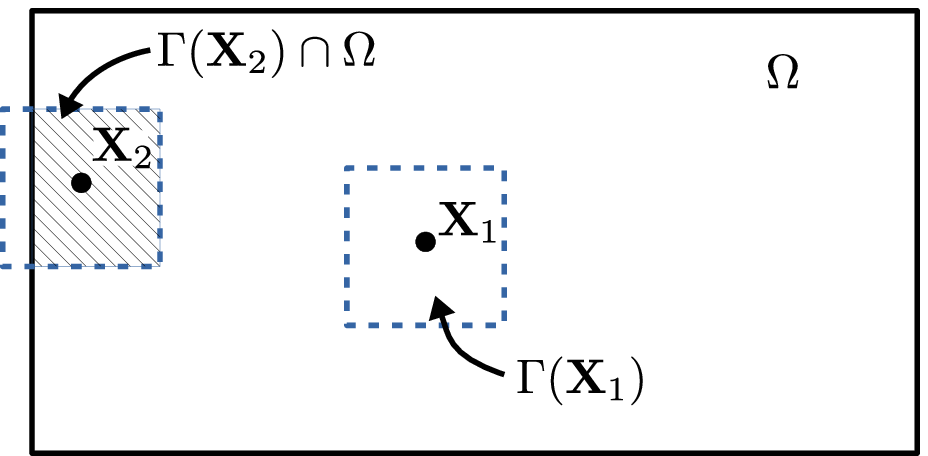}}
		\label{fig:Boundary_neighborhood}%
		\caption{Discretization of the neighborhood $\Gamma(\boldsymbol{\mathrm{X}})$, of an interior point for three different mesh refinements.}
		\label{fig:Meshed_Neighborhood}
	\end{figure}

	For the three example problems described ahead, topology optimization is performed over a rectangular domain. We discretize the domain using square elements. By the definition provided in section \ref{sec:formulation_def}, for any given $\Gamma_i \equiv \Gamma(\boldsymbol{\mathrm{X}}_i)$, element $j$ is in $\{\mathbb{N}_i\}$ if $\boldsymbol{\mathrm{X}}_j \in \Gamma_i$. Via examples we attempt to show that the nFP method leads to mesh independent solutions. We implement $\Gamma(\boldsymbol{\mathrm{X}})$ as a square region with center at $\boldsymbol{\mathrm{X}} ~\forall ~\boldsymbol{\mathrm{X}} \in \Omega$. Unlike circular neighborhoods which have been ubiquitously chosen in topology optimization, this choice of $\Gamma(\boldsymbol{\mathrm{X}})$ allows for exact discretization of neighborhoods at all points. As $\Gamma(\boldsymbol{\mathrm{X}})$ is point symmetric about $\boldsymbol{\mathrm{X}}$, $\Delta(\boldsymbol{\mathrm{X}}) = \Gamma(\boldsymbol{\mathrm{X}})$. Fig. \ref{fig:Meshed_Neighborhood} demonstrates a square $\Gamma(\boldsymbol{\mathrm{X}})$, discretized for three different mesh refinements. Fig. \ref{fig:Meshed_Neighborhood}a, b and c discretize the neighborhood using 9, 25 and 81 elements respectively. In what follows, discretization of the neighborhood is expressed using the variable $ls$, where $ls = k$ implies that the neighborhood's edge is discretized using $2k+1$ elements. Thus, fig. \ref{fig:Meshed_Neighborhood} depicts neighborhood discretization for $ls = 1, 2$ and $4$. \\
	Special consideration for the definition of neighborhood is required for points on or close to the domain boundaries. A point is said to be in the interior of the domain if $\Gamma(\boldsymbol{\mathrm{X}}) \subset \Omega$. All other points are said to be close to the boundary. For such points, neighborhood is defined as $\Gamma(\boldsymbol{\mathrm{X}}) \cap \Omega$, see fig. \ref{fig:Meshed_Neighborhood}d. Elements with centroids as interior and boundary points will be referred to as an interior and boundary elements respectively. Note that, the size of $\{\mathbb{N}_i\}$ is smaller for boundary elements.

%	\pagebreak
%	
	\section{Examples and Solutions}
\label{sec:example}
Implementing the formulation and sensitivity analysis presented in section \ref{sec:FPM}, we solve two strain energy minimization and one compliant mechanism problem. Section \ref{sec:prob_des} presents objective function and the corresponding objective gradients with respect to element densities. For both cases, we show that the objective gradients approach $\boldsymbol{0}$ as one converges to 0-1 topologies. These solutions are presented in section \ref{sec:results} along with discussions on (a) mesh independence, (b) grayness and (c) rate of convergence of solutions.

\subsection{Problem description}
\label{sec:prob_des}
This section summarizes the problem description for stiff structures and small deformation compliant mechanisms using the flexibility-stiffness formulation \citep{saxena2000optimal}.

\subsubsection{Compliance Minimization Problem}
For compliance minimization problems, the objective $\mathbb{I}(\boldsymbol{\rho})$ is the strain energy stored, that is,
\begin{eqnarray}
\mathbb{I}(\boldsymbol{\rho}) = \lambda\frac{1}{2}\boldsymbol{\mathrm{u}}^T \boldsymbol{\mathrm{K}}(\boldsymbol{\rho}) \boldsymbol{\mathrm{u}}.
\label{eqn:com_min_obj}
\end{eqnarray}
where $\lambda$ is a scalar parameter. Correspondingly, gradient of the objective with respect to element densities is given as,
\begin{eqnarray}
\frac{\partial \mathbb{I}(\boldsymbol{\rho})}{\partial \rho_i} =  -\lambda\frac{1}{2} \boldsymbol{\mathrm{u}}^T \frac{\partial \boldsymbol{\mathrm{K}}(\boldsymbol{\rho})}{\partial \rho_i} \boldsymbol{\mathrm{u}} =
- \lambda\frac{\eta (1-\rho_{min})}{2} \rho^{\eta -1}_i \boldsymbol{\mathrm{u}}_i^T \boldsymbol{\mathrm{K}}_0\boldsymbol{\mathrm{u}}_i
\label{eqn:comp_mini_grad}
\end{eqnarray} 
where $\frac{\partial \boldsymbol{\mathrm{K}(\boldsymbol{\rho})}}{\partial \rho_i}$ is evaluated using eqn. \ref{eqn:SIMP} and $\boldsymbol{\mathrm{u}}_i$ is the local displacement vector for element $i$. Combining eqns. \ref{eqn:gradient} and \ref{eqn:comp_mini_grad}, gradient of $\mathbb{I}(\boldsymbol{\rho})$ with respect to design variables can be expressed as,
\begin{eqnarray}
\frac{\partial \mathbb{I}}{\partial \beta_j} = \sum\limits_{i\in \{\mathbb{N}_j\}}\lambda\frac{\eta (1-\rho_{min})}{2} \rho^{\eta -1}_i (1-\rho_i) \boldsymbol{\mathrm{u}}_i^T \boldsymbol{\mathrm{K}}_0\boldsymbol{\mathrm{u}}_i \dfrac{A(\Omega_j)}{A(\Delta_i)}.
\label{eqn:comp_min_design_var_grad}
\end{eqnarray}
The above expression is $0$ when $\rho_i = 0 ~\text{or}~ 1 ~\forall~ i \in \{\mathbb{N}_j\}$. This implies that the gradient diminishes as one approaches black and white topologies. For the density evaluation method discussed in sec. \ref{sec:density_eval}, $\rho_i = 1$ is not possible and thus points with precisely $\boldsymbol{0}$ gradients are not part of the solution space.

\begin{figure}[!htb]
	\begin{minipage}{0.45\textwidth}
		\centering
		\vspace{4mm}
		\includegraphics[width=0.7\textwidth]{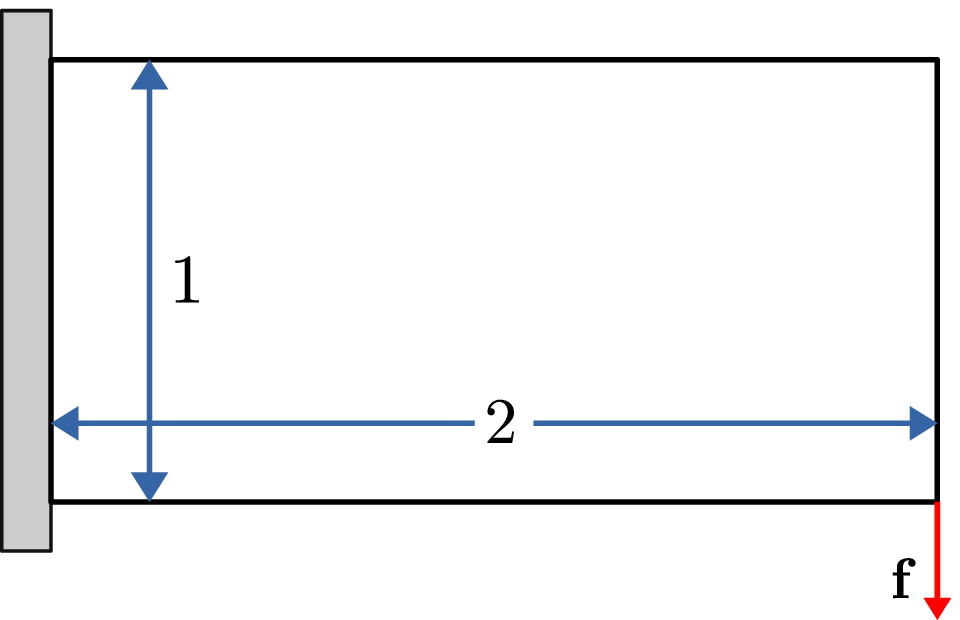}
		\caption{Problem description for stiff cantilever beam design. }
		\label{fig:Prob_description_Canti}
	\end{minipage}\hfill
	\begin{minipage}{0.5\textwidth}
		\centering
%		\vspace{-2mm}
		\includegraphics[width=0.95\textwidth]{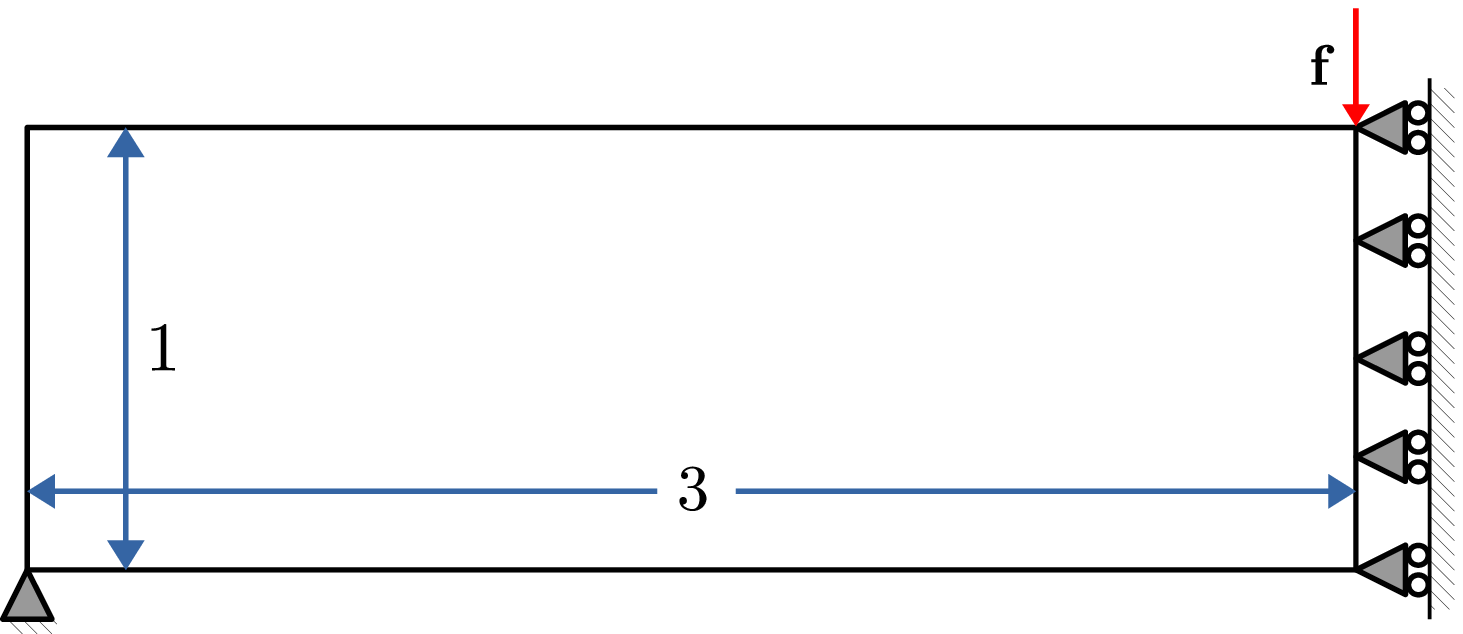}
		\vspace{3mm}
		\caption{Problem description for stiff MBB beam design.}
		\label{fig:Prob_description_MBB}
	\end{minipage}
\end{figure}

Herein, we solve two compliance minimization problems namely, cantilever beam design (fig. \ref{fig:Prob_description_Canti}) and MBB beam design (fig. \ref{fig:Prob_description_MBB}). The cantilever beam problem is solved over a rectangular domain with the left boundary fixed and a vertical downward force $\boldsymbol{\mathrm{f}}$ applied at the right bottom corner, see fig. \ref{fig:Prob_description_Canti}. The MBB beam problem is also solved over a rectangular domain with the left bottom corner fixed, the right boundary fixed along the horizontal while free to move along the vertical and a vertical, downward force $\boldsymbol{\mathrm{f}}$ applied at the right top corner of the domain, see fig. \ref{fig:Prob_description_MBB}.

\subsubsection{Compliant Mechanism Problem}
In compliant mechanism design, the aim is to develop continua which provide maximum desired deflection at a given location in a certain direction for specified traction and displacement boundary conditions over the design domain. Objective function, $\mathbb{I}(\boldsymbol{\rho})$ in eqn. \ref{eqn:2}, for compliant mechanism problems is given as
\begin{eqnarray}
\label{eqn:compliant_mech}
\mathbb{I}(\boldsymbol{\rho}) = -\lambda \frac{\boldsymbol{\mathrm{v}}^T \boldsymbol{\mathrm{K}} \boldsymbol{\mathrm{u}}}{\boldsymbol{\mathrm{u}}^T \boldsymbol{\mathrm{K}} \boldsymbol{\mathrm{u}}}
\end{eqnarray}
where $\lambda$ is a scalar parameter, $\boldsymbol{\mathrm{v}}$ is the displacement vector obtained in response to a dummy force vector, $\boldsymbol{\mathrm{F}}_d$ \citep{yin2003design}, that is, $\boldsymbol{\mathrm{K}} \boldsymbol{\mathrm{v}} = \boldsymbol{\mathrm{F}}_d$. The dummy force vector has a unit magnitude at output degrees of freedom, that is, nodes at which the deflection is captured, in the direction of desired deflection. Hence, numerator of the objective in Eqn. \ref{eqn:compliant_mech}, $\boldsymbol{\mathrm{v}}^T \boldsymbol{\mathrm{K}} \boldsymbol{\mathrm{u}} = \boldsymbol{\mathrm{F}}_d^T \boldsymbol{\mathrm{u}}$ (using symmetry of $\boldsymbol{\mathrm{K}}$), is the same as displacement at the output nodes in the desired direction. Thus the muti-criteria objective attempts to minimize strain energy (a measure of internal strength) while maximizing the output displacements (a measure of flexibility) \citep{saxena2000optimal}. Force transfer is ensured by adding an artificial stiffness thereby replacing the stiffness matrix, $\boldsymbol{\mathrm{K}}$, in Eqn. \ref{eqn:compliant_mech} by $\boldsymbol{\mathrm{K}} = \boldsymbol{\mathrm{K}} + \boldsymbol{\mathrm{K}}_a$ where $\boldsymbol{\mathrm{K}}_a$ is the artificial stiffness matrix exhibiting non-zero components only on diagonal elements corresponding to output degrees of freedom.

Gradient of the objective with respect to element densities is given as
\begin{eqnarray}
\frac{\partial \mathbb{I}(\boldsymbol{\rho})}{\partial \rho_i} = \frac{1}{\boldsymbol{\mathrm{u}}^T \boldsymbol{\mathrm{K}} \boldsymbol{\mathrm{u}}} \left\{ \lambda\left( \boldsymbol{\mathrm{v}}^T \frac{\partial \boldsymbol{\mathrm{K}}}{\partial \rho_i} \boldsymbol{\mathrm{u}} \right) + \mathbb{I}(\boldsymbol{\rho}) \left( \boldsymbol{\mathrm{u}}^T \frac{\partial \boldsymbol{\mathrm{K}}}{\partial \rho_i} \boldsymbol{\mathrm{u}} \right) \right\}
\label{eqn:CM_density_grad}
\end{eqnarray}
Evaluation of $\frac{\partial \boldsymbol{\mathrm{K}}}{\partial \rho_i}$ is same as in eqn. \ref{eqn:comp_mini_grad}. Note that $\boldsymbol{\mathrm{K}}_a$ is constant and thus, it does not contribute to the evaluation of $\frac{\partial \boldsymbol{\mathrm{K}}}{\partial \rho_i}$. Combining eqn. \ref{eqn:CM_density_grad} and eqn. \ref{eqn:gradient} it can be realized that, similar to the compliance minimization problem, gradient for the compliant mechanism problem is also identically $\boldsymbol{0}$ for a pure 0-1 topology.

\begin{figure}[!htb]
	\centering
	\includegraphics[width=0.35\textwidth]{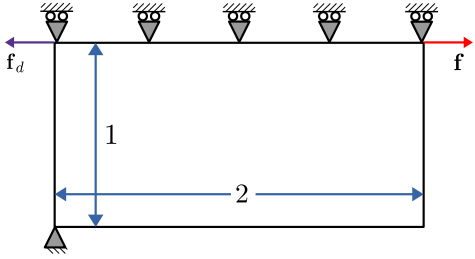}
	\caption{Problem description for displacement inverter.}
	\label{fig:Prob_description_disp_invert}
\end{figure}

For the displacement inverter problem, solved herein, the objective is to develop a continuum such that displacement at the output node is in opposite direction to displacement at the input nodes. Design domain and the corresponding boundary conditions are depicted in fig. \ref{fig:Prob_description_disp_invert}.

\subsection{Results}
\label{sec:results}
To demonstrate that the proposed methods leads to mesh independent solutions, we solve the above problems for various mesh sizes while maintaining the shape of $\Gamma(\boldsymbol{\mathrm{X}})$. Results of each example are analyzed under the following three categories: (a) mesh independence, (b) grayness and (c) rate of convergence of solutions. To measure grayness of a solution, we implement the grayness function \citep{Singh2020} given by,
\begin{eqnarray}
g(\boldsymbol{\rho}) = \sum\limits_{i = 1}^{ne} \frac{4\rho_i(1-\rho_i)}{ne} 
\end{eqnarray}
where $ne$ is the total number of elements. Note that $g(\boldsymbol{\rho}) = 1$ for $\rho_i = 0.5$ throughout while $g(\boldsymbol{\rho}) = 0$ for purely black and white solutions. The formulation described in eqn. \ref{eqn:2} is non-convex, thus the solution depends on the initial guess. To this end, for the initial guess, we set $\beta_i = \ln(0.3)~ \forall ~i \in \{1,2,3 \hdots ne\}$. This gives a uniform density distribution of $\rho_i = 0.7$ throughout the design domain. Lower bound on the design parameters is kept as $\beta_{lb} = -10 \times d_n$ where $d_n$ is the size of $\{\mathbb{D}_i\}$ for an interior element $i$. Parameters for the SIMP material model are maintained at $\eta = 3$ and $\rho_{min} = 10^{-4}$ for all solutions presented. Evaluating the objective for a given continua requires a linear plane strain FE analysis, which is conducted using the following material properties: $E = 2\times 10^{4}$ and $\nu = 0.3$. For all the solutions presented ahead the grayness measure, $g(\boldsymbol{\rho})$, along with various other parameters is mentioned in captions below the figure. All problems are solved using the inbuilt MATLAB function \textit{fmincon} \citep{MATLAB:2010}.

\subsubsection{Cantilever beam problem}
\label{sec:canti_results}
\begin{figure}[!htb]
	\centering
	\subcaptionbox{Solution for mesh size: $100\times 50,~ vf = 0.35,~ ls = 2~\text{and}~ \lambda = 10^4.~ g(\boldsymbol{\rho}) = 8.8\times 10^{-3} $. }{\includegraphics[width=0.4\textwidth]{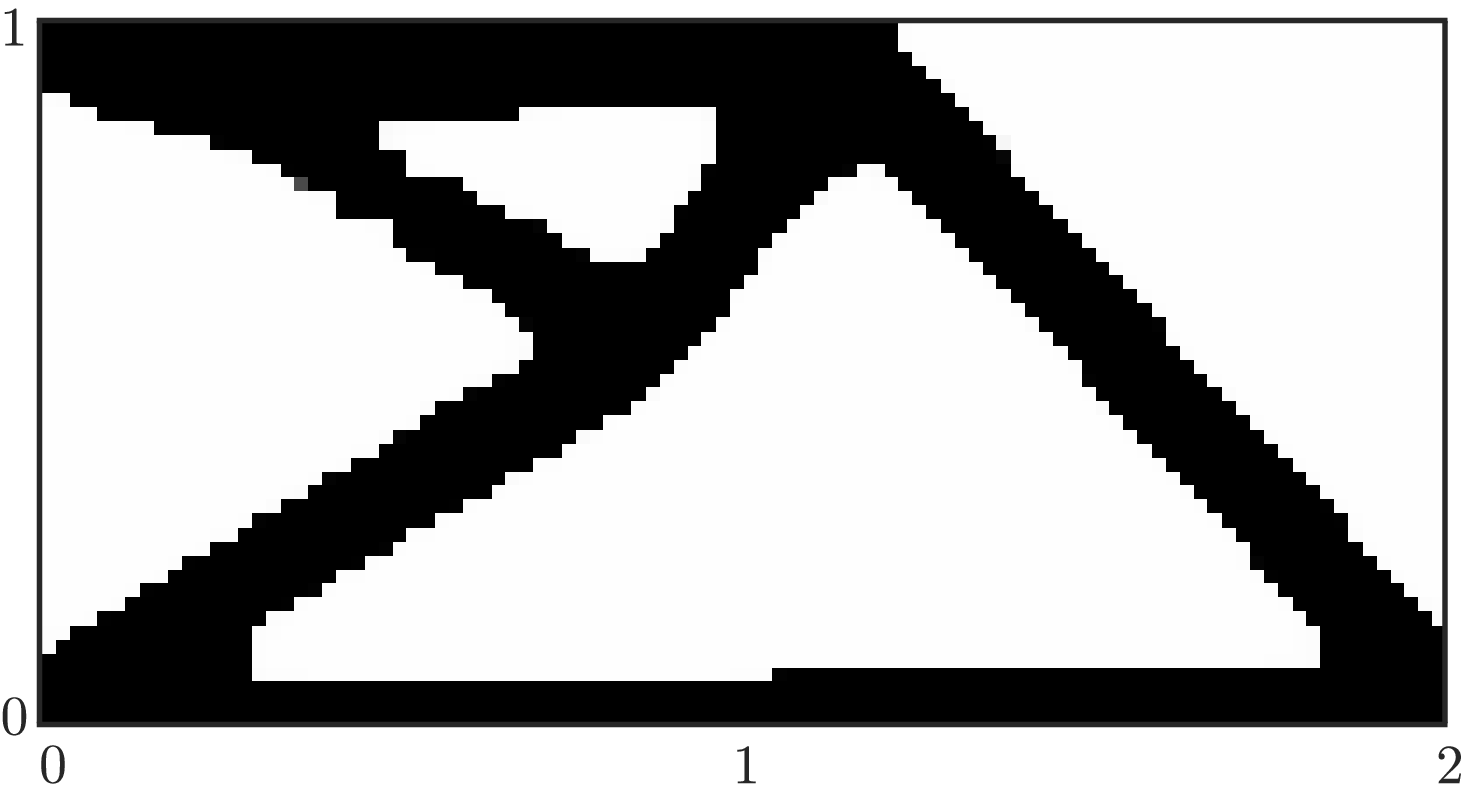}}
	\label{fig:Sol_Canti_100x50_vf_35_ls_2}%
	\hspace{0.4cm}
	%	\\[12pt]
	\subcaptionbox{Solution for mesh size: $140\times 70,~ vf = 0.35,~ ls = 3~\text{and}~ \lambda = 5\times10^3.~ g(\boldsymbol{\rho}) = 1.04\times 10^{-2} $. }{\includegraphics[width=0.4\textwidth]{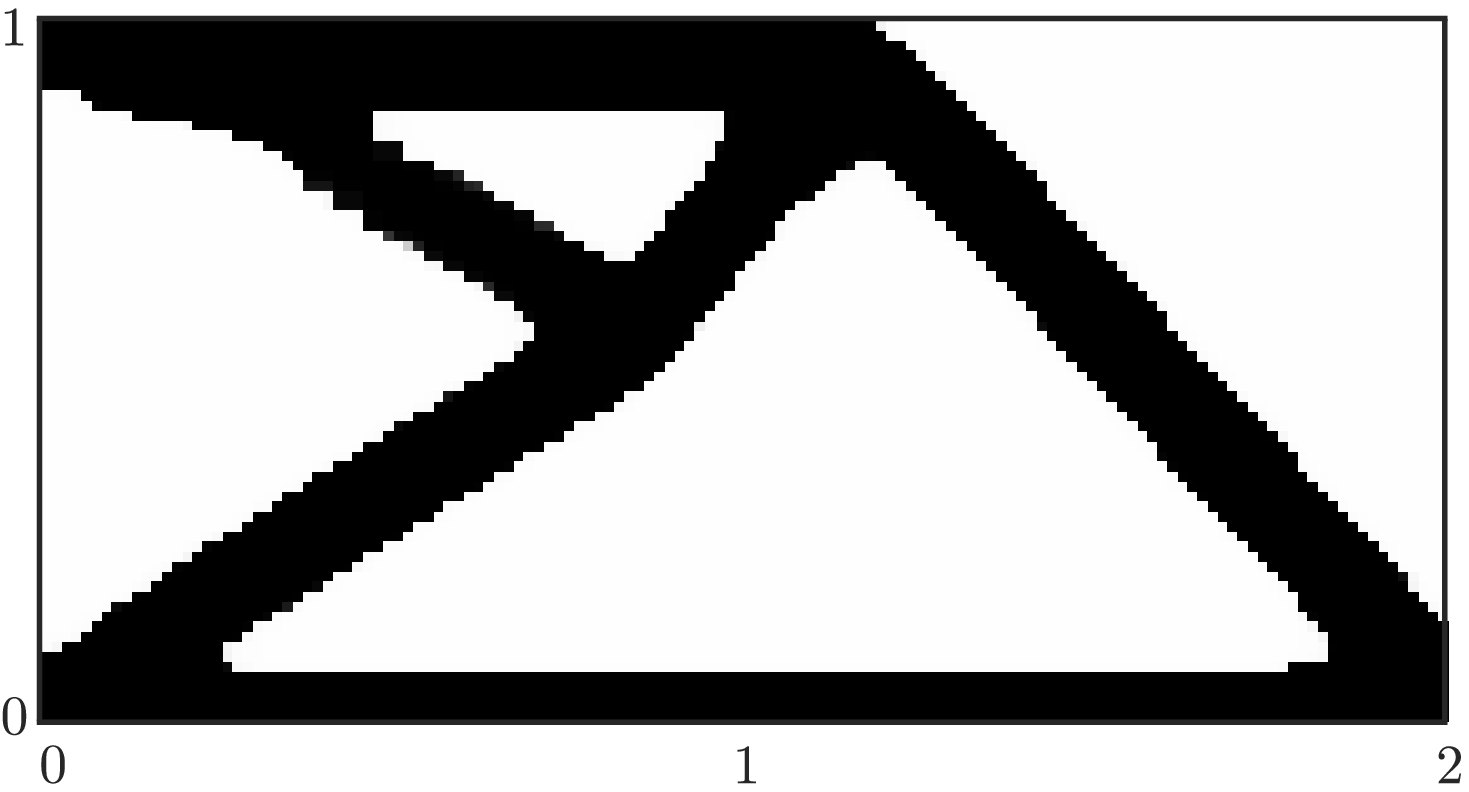}}
	\label{fig:Sol_Canti_140x70_vf_35_ls_3}%
	\\[12pt]
	\subcaptionbox{Solution for mesh size: $180\times 90,~ vf = 0.35,~ ls = 4~\text{and}~ \lambda = 2 \times 10^4.~ g(\boldsymbol{\rho}) = 8.5 \times 10^{-3} $.}{\includegraphics[width=0.4\textwidth]{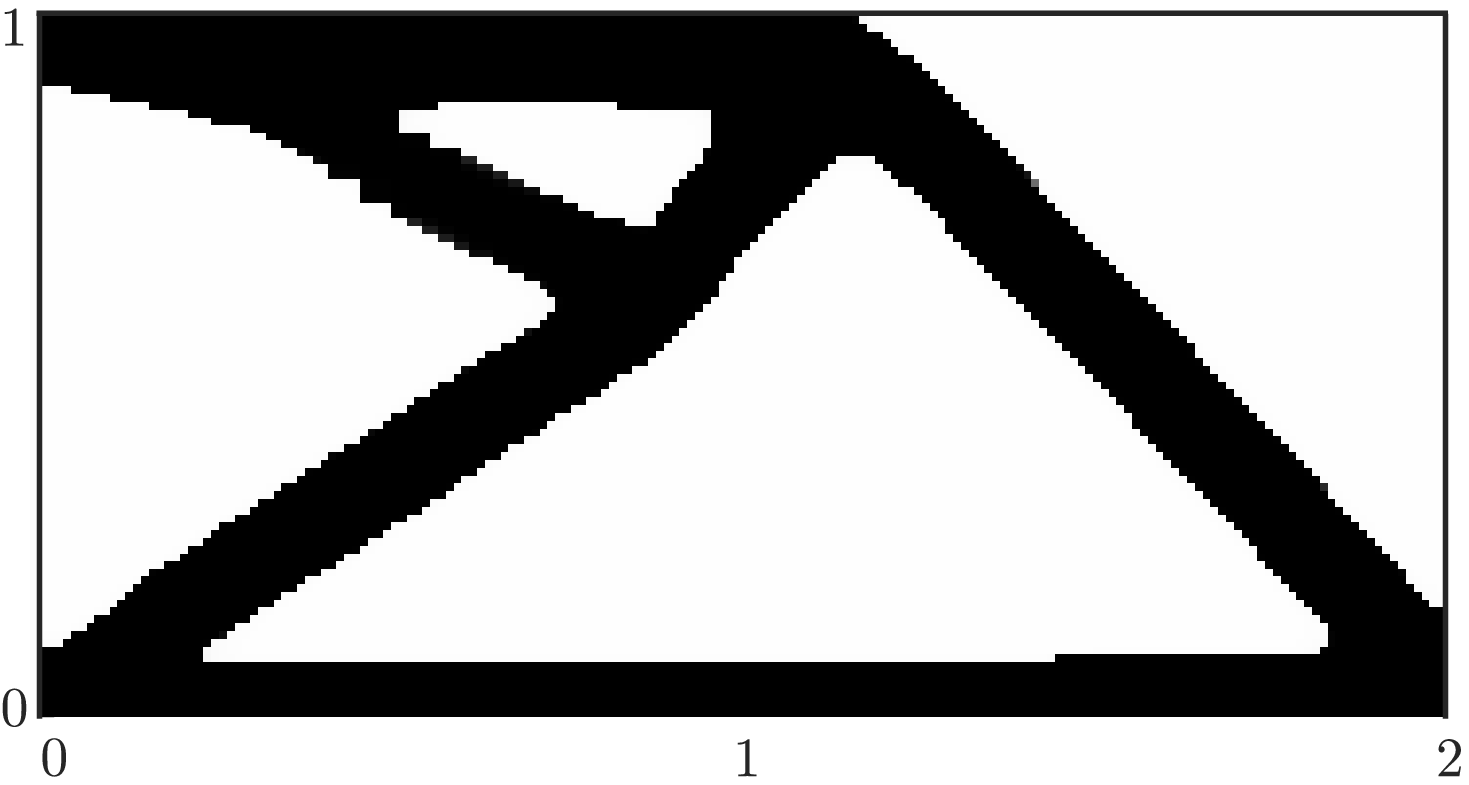}}
	\label{fig:Sol_Canti_180x90_vf_35_ls_4}%
	\hspace{0.4cm}
	\subcaptionbox{Solution for mesh size: $120\times 60,~ vf = 0.35,~ ls = 1~\text{and}~ \lambda = 10^3.~ g(\boldsymbol{\rho}) = 8.0 \times 10^{-3} $.}{\includegraphics[width=0.4\textwidth]{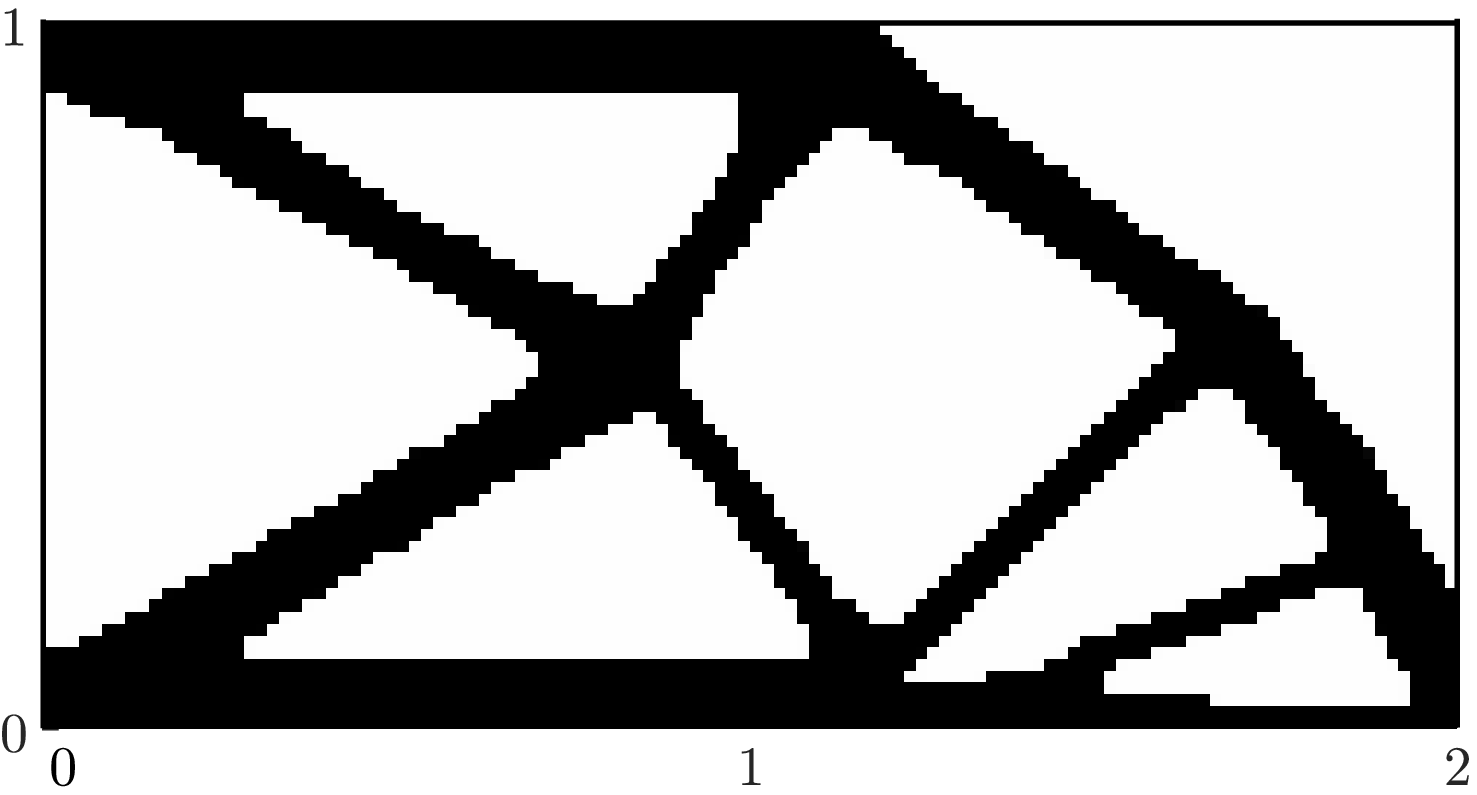}}
	\label{fig:Sol_Canti_120x60_vf_35_ls_1}
	\\[12pt] 
	%	\\[12pt]
	\subcaptionbox{Solution for mesh size: $120\times 60,~ vf = 0.35,~ ls = 2~\text{and}~ \lambda = 10^3.~ g(\boldsymbol{\rho}) = 1.7 \times 10^{-3} $.}{\includegraphics[width=0.4\textwidth]{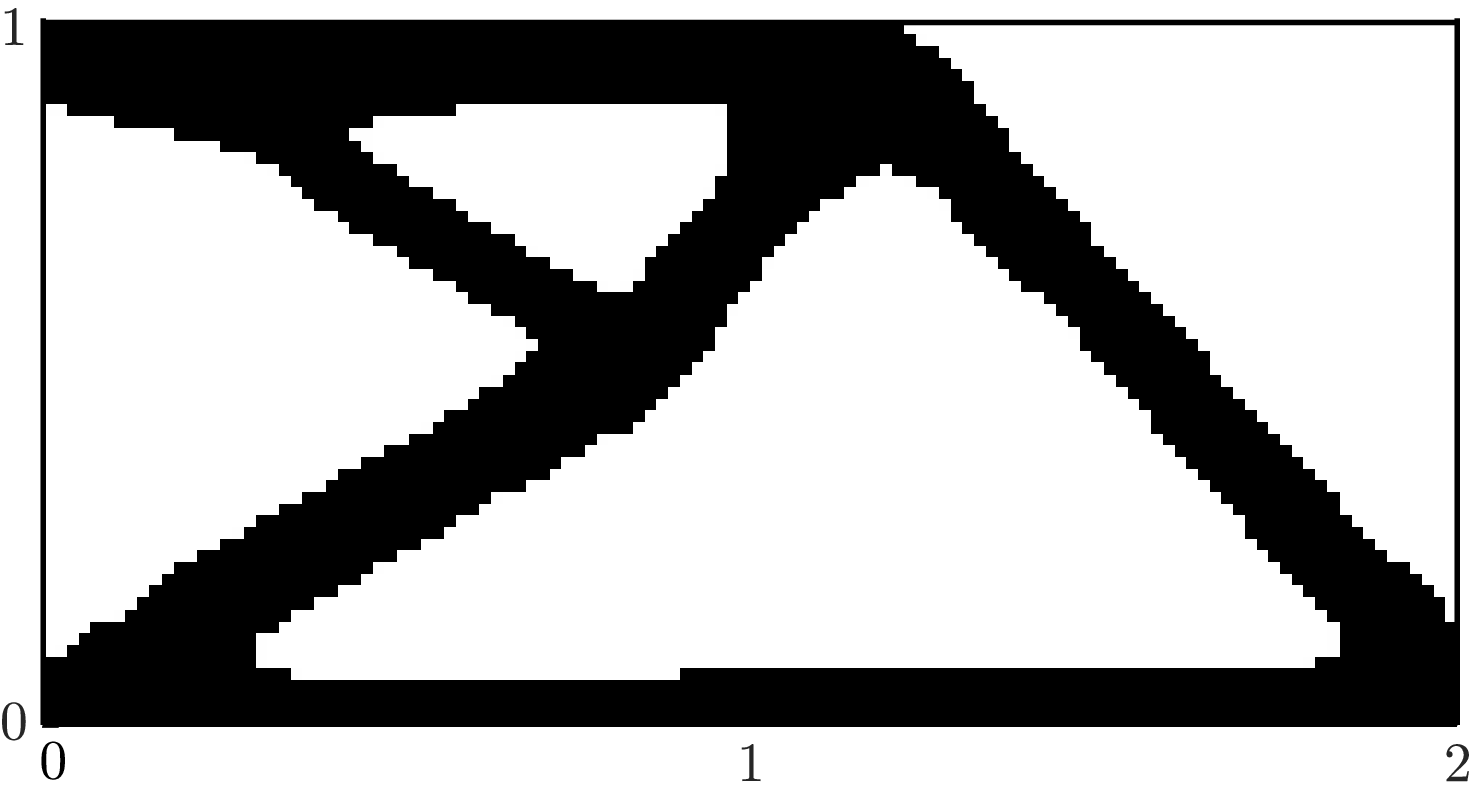}}
	\label{fig:Sol_Canti_120x60_vf_35_ls_2}%
	\hspace{0.4cm}
	\subcaptionbox{Solution for mesh size: $120\times 60,~ vf = 0.35,~ ls = 3~\text{and}~ \lambda = 10^3.~ g(\boldsymbol{\rho}) = 9.2 \times 10^{-3} $.}{\includegraphics[width=0.4\textwidth]{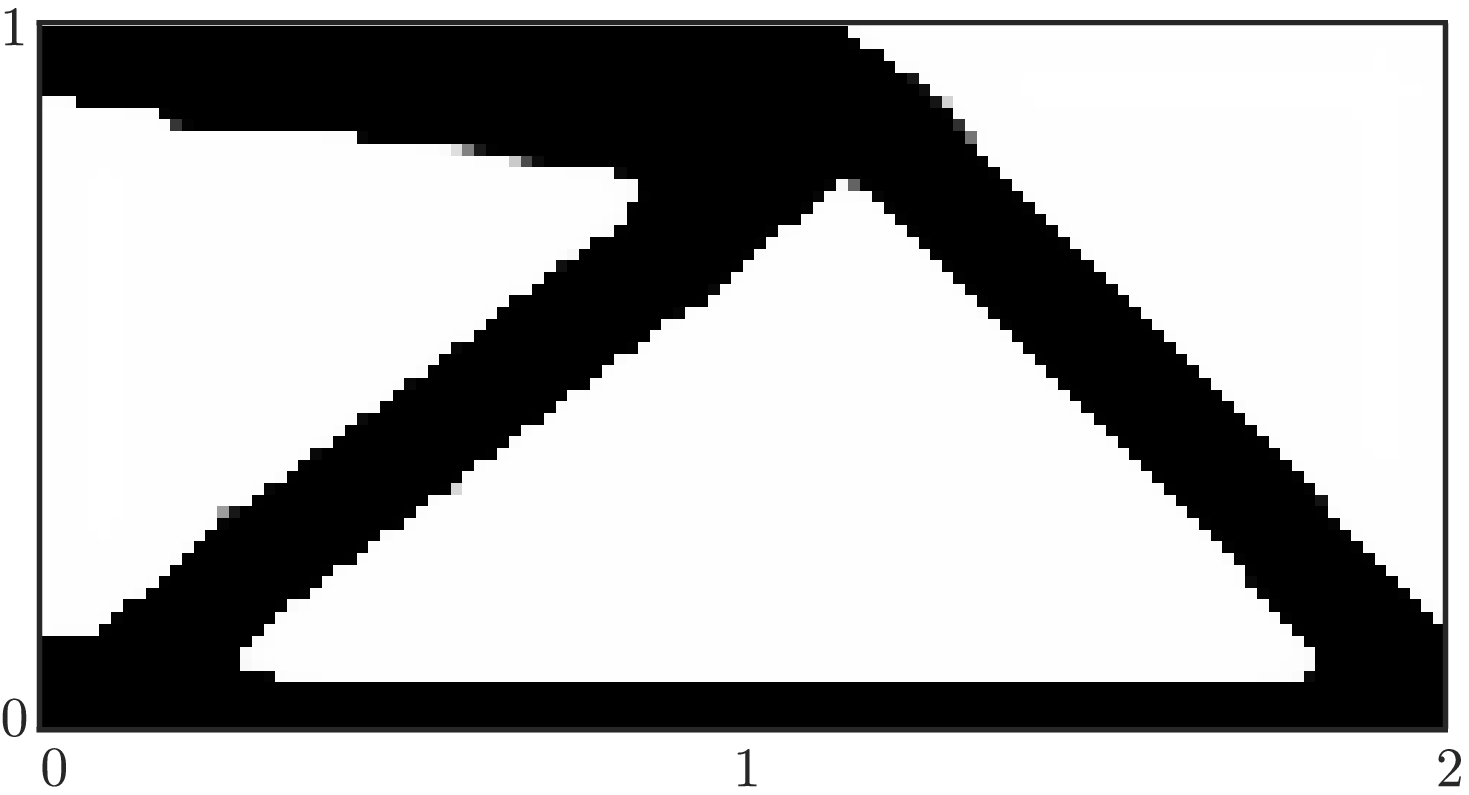}}
	\label{fig:Sol_Canti_120x60_vf_35_ls_3}%
	\\[12pt]
%	\hspace{0.4cm}
	%	\\
	\subcaptionbox{Convergence history for solution in fig. \ref{fig:Sol_Canti}d. }{\includegraphics[width=0.45\textwidth]{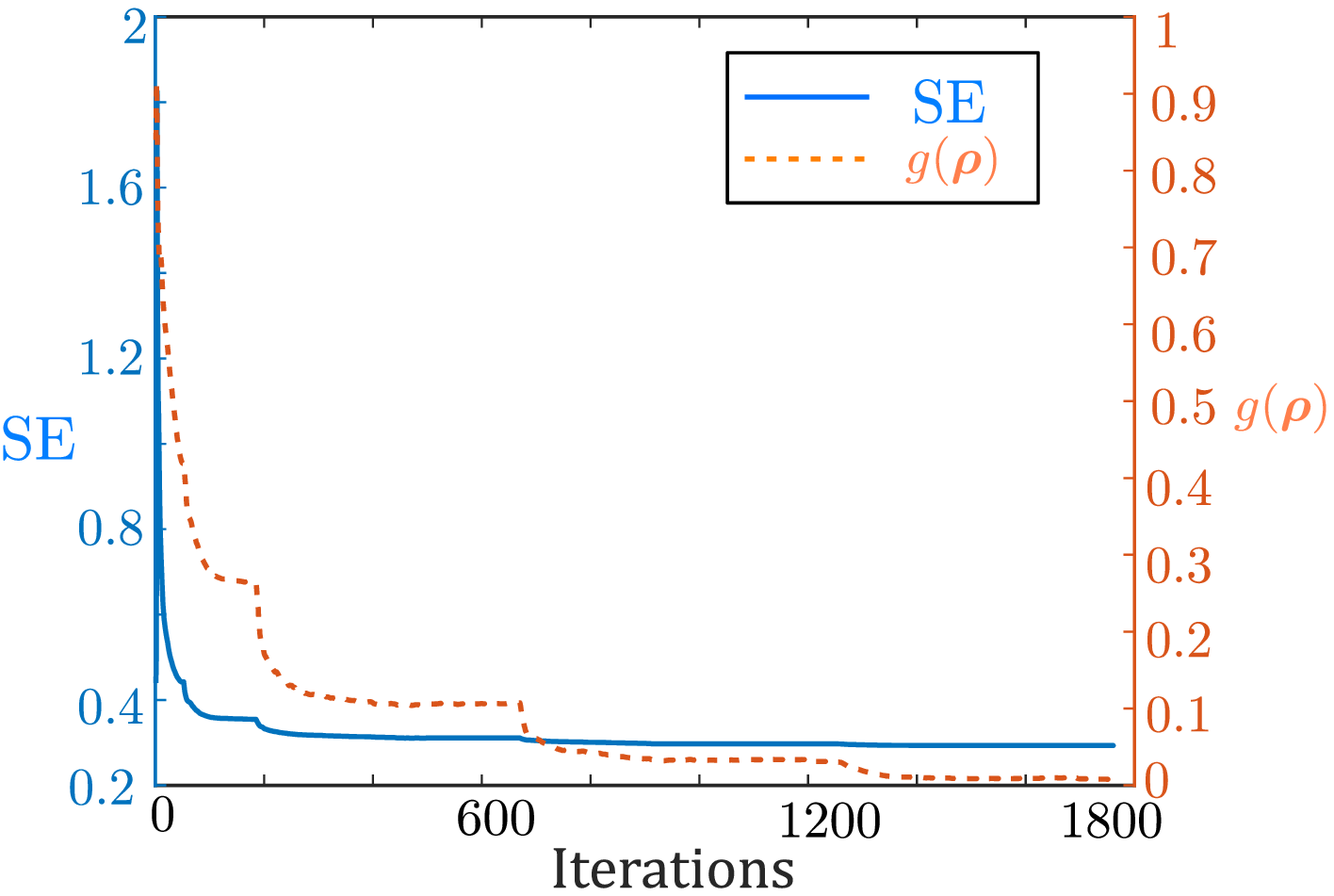}}
	\label{fig:Sol_Canti_120x60_vf_35_ls_1_convergence}
	\caption{Solutions to the cantilever beam problem.}
	\label{fig:Sol_Canti}
\end{figure}

We present solutions to the cantilever beam problem for six different cases (fig. \ref{fig:Sol_Canti}). Volume fraction for all solutions is, $vf = 0.35$. Fig. \ref{fig:Sol_Canti}a, b and c discretize the domain into $100 \times 50$, $140 \times 70$ and $180 \times 90$ elements, and implement $ls=2,3$ and $4$ respectively, thus presenting solutions for varied domain discretization while maintaining $\Gamma(\boldsymbol{\mathrm{X}})$. Fig. \ref{fig:Sol_Canti}d, e and f, present solutions for a discretization of $120 \times 60$ and varied $ls$, specifically $ls = 1,2$ and $3$ respectively. Solutions in fig. \ref{fig:Sol_Canti}a,b and c have identical density distribution suggesting mesh independent solutions can be achieved with the presented formulation. Different topologies are observed in fig. \ref{fig:Sol_Canti}d, e and f suggesting different topologies can be captured by varying the size of $\Gamma(\boldsymbol{\mathrm{X}})$. 
%The $g(\boldsymbol{\rho})$ for each solution is mentioned in the caption, along with various other parameters implemented to obtain the solution.
For all solutions $g(\boldsymbol{\rho})$ is close to or below $10^{-2}$ or $1\%$. It is evident from the solutions, that the proposed method can avoid any transition region. On careful inspection some gray cells are found along edges of the structures. Potential reasons for their presence are discussed in section \ref{sec:discussion}. Fig. \ref{fig:Sol_Canti}g presents convergence history for the solution in fig. \ref{fig:Sol_Canti}d. The objective value converges around 600 iterations while the grayness function takes much longer to converge. A similar pattern was observed for other solutions as well. This slow rate of drop is attributed to small magnitude of gradients close to pure black and white topologies. 

\subsubsection{MBB beam problem}
\label{sec:MBB_results}
\begin{figure}[!htb]
	\centering
	\subcaptionbox{Solution for mesh size: $105\times 35,~ vf = 0.35,~ ls = 2~\text{and}~ \lambda = 5 \times 10^3.~ g(\boldsymbol{\rho}) = 6.2 \times 10^{-3}$. }{\includegraphics[width=0.45\textwidth]{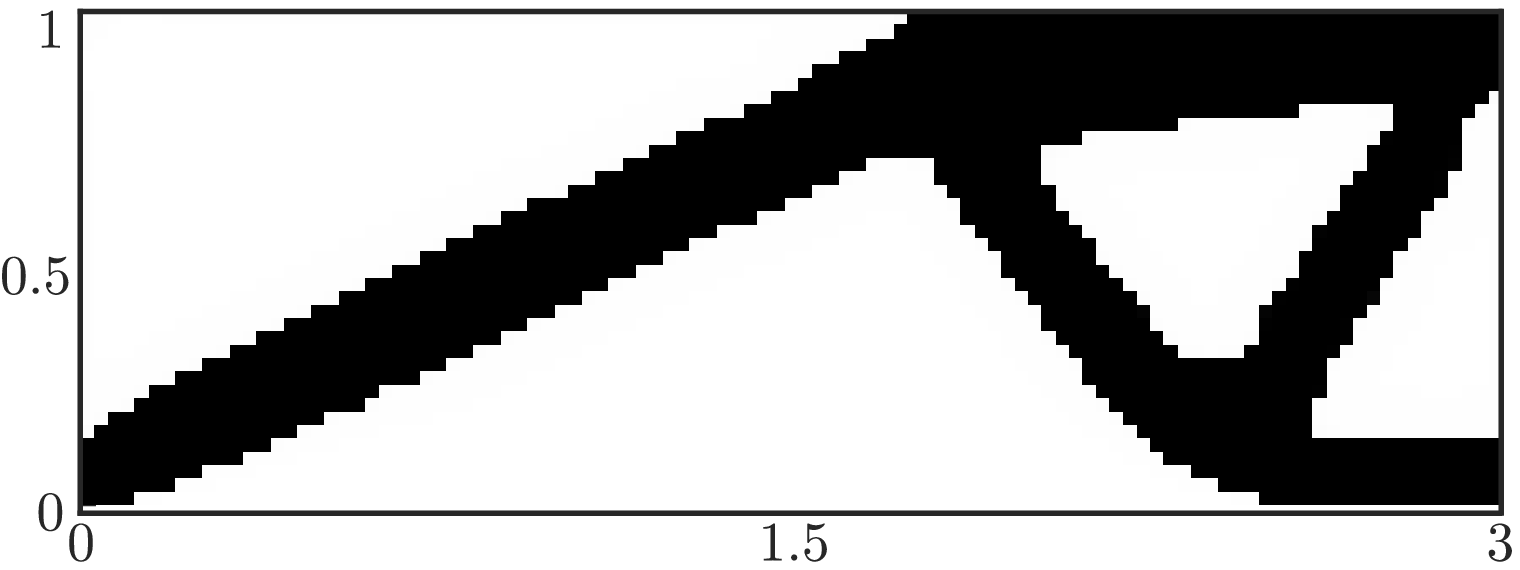}}
	\label{fig:Sol_MBB_105x35_vf_35_ls_2}%
	\hspace{0.4cm}
%	\\[12pt]
	\subcaptionbox{Solution for mesh size: $147\times 49,~ vf = 0.35,~ ls = 3~\text{and}~ \lambda = 5 \times 10^3.~ g(\boldsymbol{\rho}) = 9.7 \times 10^{-3} $.}{\includegraphics[width=0.45\textwidth]{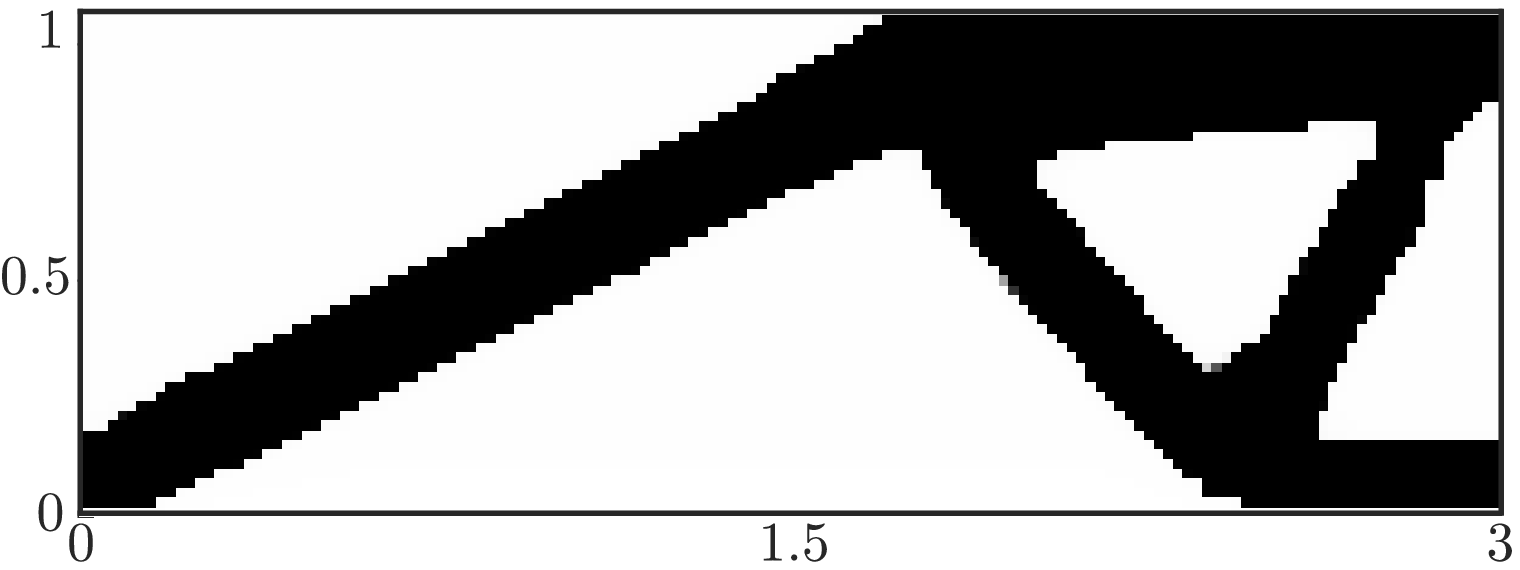}}
	\label{fig:Sol_MBB_147x49_vf_35_ls_3}%
%	\hspace{0.7cm}
	\\[12pt]
	\subcaptionbox{Solution for mesh size: $189\times 63,~ vf = 0.35,~ ls = 4~\text{and}~ \lambda = 10^4.~ g(\boldsymbol{\rho}) = 1.01 \times 10^{-2} $.}{\includegraphics[width=0.45\textwidth]{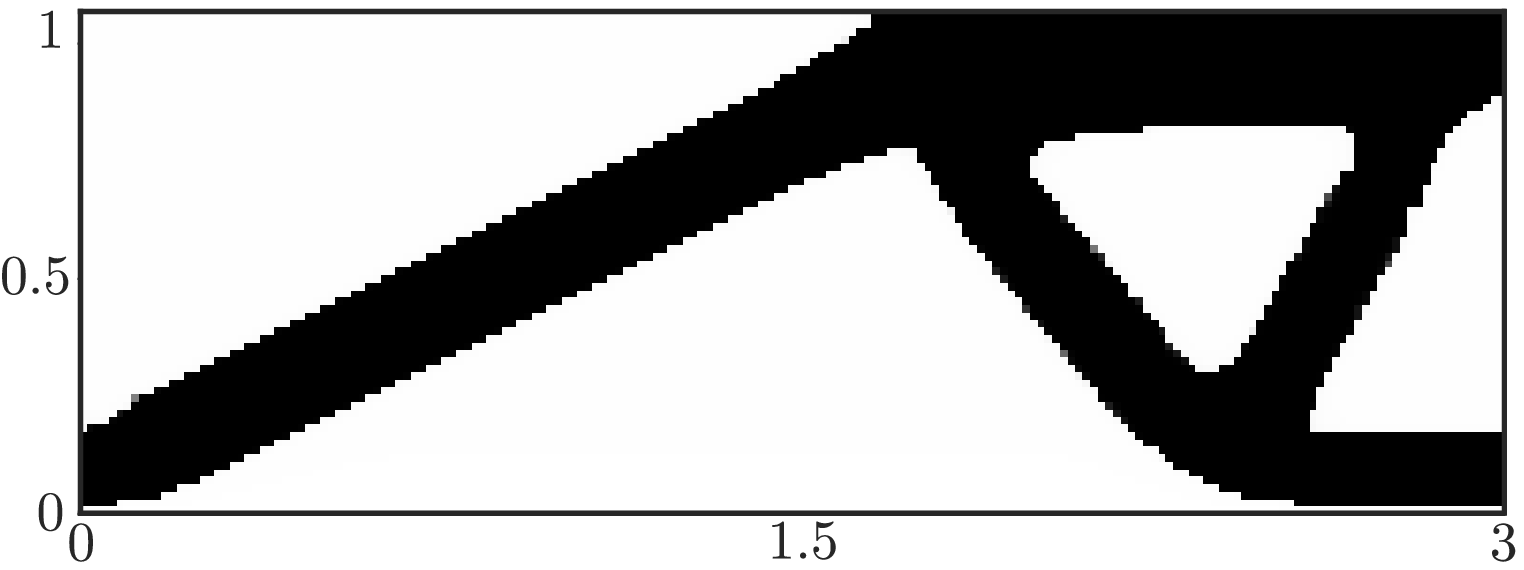}}
	\label{fig:Sol_MBB_189x63_vf_35_ls_4}%
	\hspace{0.4cm}
	\subcaptionbox{Solution for mesh size: $120\times 40,~ vf = 0.35,~ ls = 2~\text{and}~ \lambda = 10^3.~ g(\boldsymbol{\rho}) = 7.2 \times 10^{-3} $.}{\includegraphics[width=0.45\textwidth]{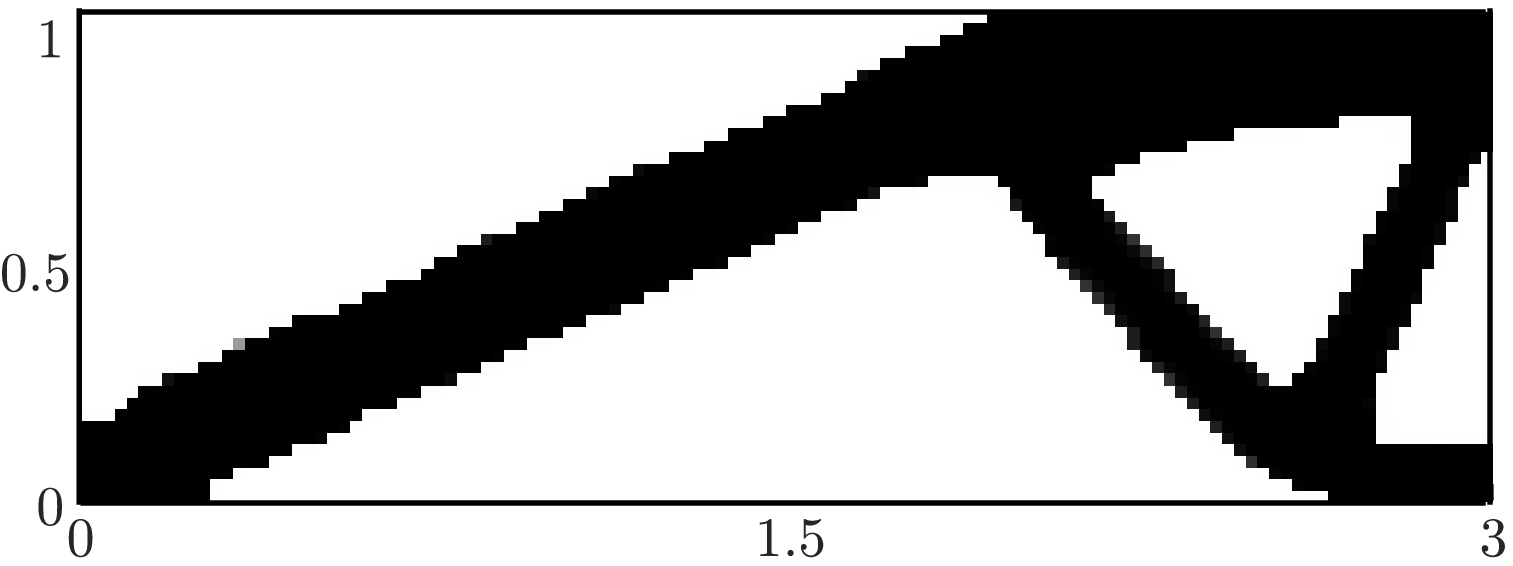}}
	\label{fig:Sol_MBB_120x40_vf_35_ls_2}%
%	\hspace{0.4cm} 
	\\[12pt]
	\subcaptionbox{Solution for mesh size: $120\times 40,~ vf = 0.25,~ ls = 2~\text{and}~ \lambda = 10^3.~ g(\boldsymbol{\rho}) = 6.4 \times 10^{-3} $.}{\includegraphics[width=0.45\textwidth]{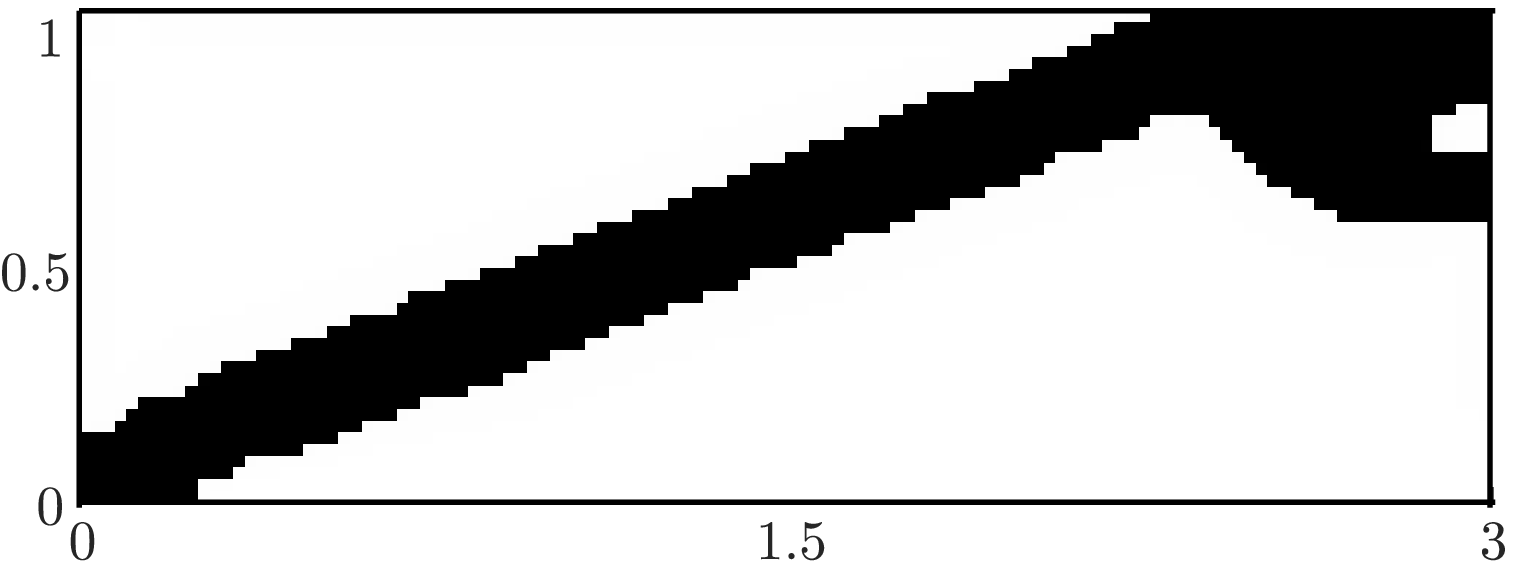}}
	\label{fig:Sol_MBB_120x40_vf_25_ls_2}%
	\hspace{0.4cm}
%	\\[12pt]
	\subcaptionbox{Solution for mesh size: $120\times 40,~ vf = 0.18,~ ls = 2~\text{and}~ \lambda = 10^3.~ g(\boldsymbol{\rho}) = 4.1 \times 10^{-3} $.}{\includegraphics[width=0.45\textwidth]{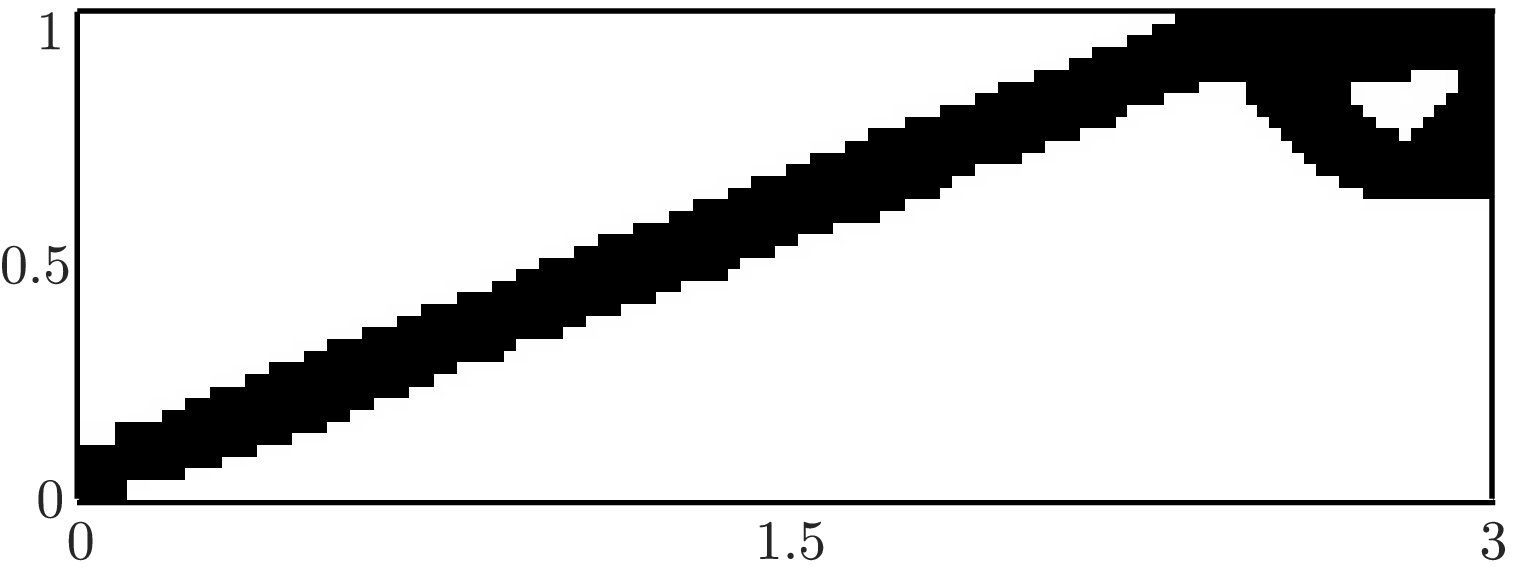}}
	\label{fig:Sol_MBB_120x40_vf_18_ls_2}
%	\hspace{0.4cm}
	\\[12pt]
	\subcaptionbox{Convergence history for solution in fig. \ref{fig:Sol_MBB}d. }{\includegraphics[width=0.45\textwidth]{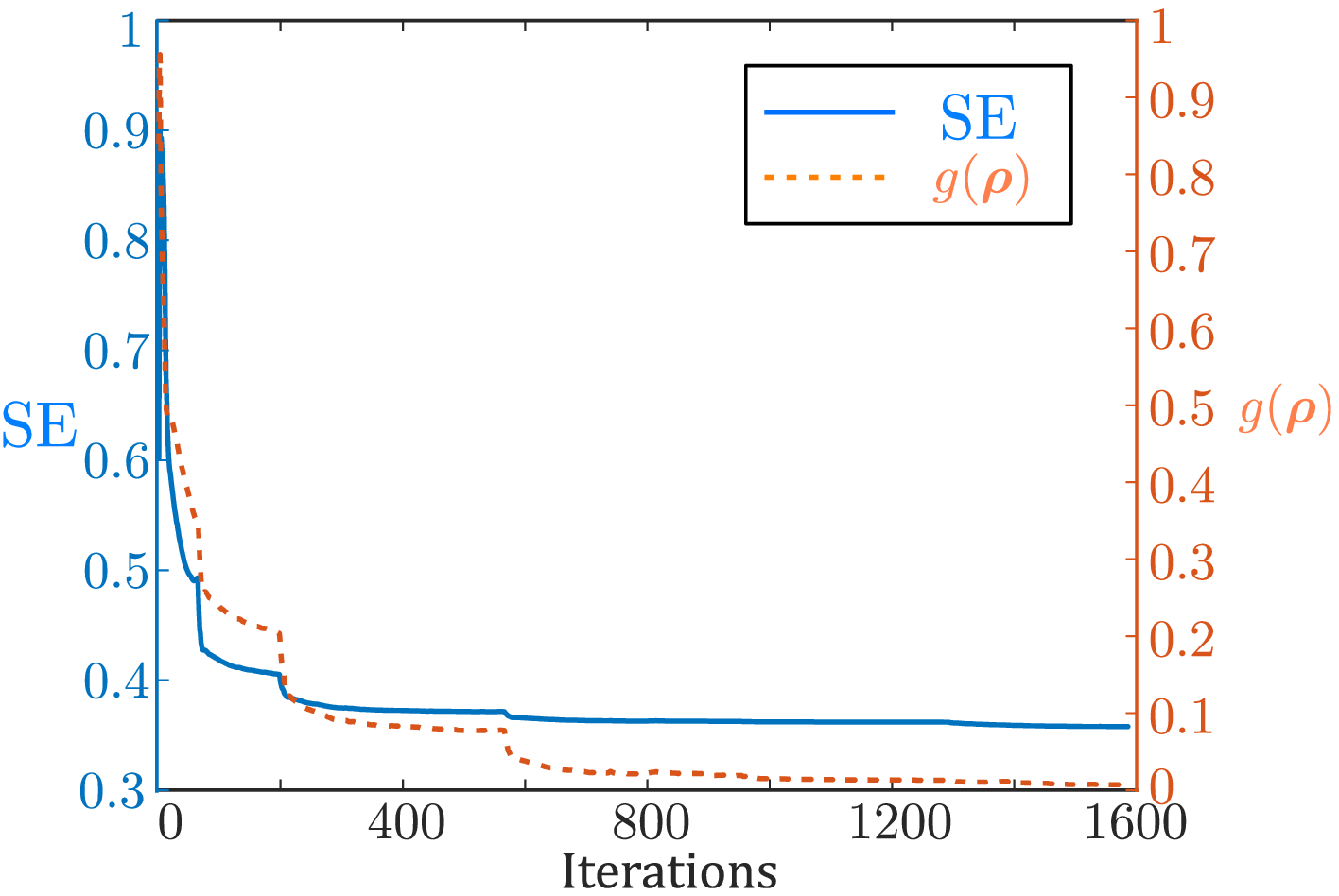}}
	\label{fig:Sol_MBB_120x40_vf_35_ls_2_convergence}
	\caption{Solutions to the MBB beam problem.}
	\label{fig:Sol_MBB}
\end{figure}

As the cantilever beam problem, we present solutions to the MBB beam problem for six different cases (fig. \ref{fig:Sol_MBB}). Fig. \ref{fig:Sol_MBB}a, b and c discretize the domain into $105 \times 35$, $147 \times 49$ and $189 \times 63$ elements and implement $ls=2,3$ and $4$ respectively, thus presenting solutions for varied domain discretization while maintaining the shape and size of $\Gamma(\boldsymbol{\mathrm{X}})$. The volume fraction for these solutions is $vf = 0.35$. Fig \ref{fig:Sol_MBB}d, e and f, present solutions for a discretization of $120 \times 40$ and $ls = 2$ for $vf = 0.35,~ 0.25$ and $0.18$ respectively. Solutions in fig. \ref{fig:Sol_MBB}a, b and c have identical density distribution suggesting mesh independence of the solution. Varied topologies are observed in fig. \ref{fig:Sol_MBB}d, e and f showing that different topologies can be captured by varying volume fraction. 
%The grayness measure for each solution is mentioned in captions below the figure, along with various parameters. 
$g(\boldsymbol{\rho})$ for all solutions is very close to or below $10^{-2}$ or $1\%$. For fig. \ref{fig:Sol_MBB}b and c some gray cells can be found at some places along the structures' boundaries. Fig. \ref{fig:Sol_MBB}g presents the convergence history for the solution in fig. \ref{fig:Sol_MBB}d. Similar to the cantilever beam problem, it is noted that compared to the objective, $g(\boldsymbol{\rho})$ takes much longer to converge and drop below an acceptable value.

Obtaining the solutions presented in fig. \ref{fig:Sol_MBB} requires a special boundary consideration. We introduce $ls$ rows of dummy elements along the bottom edge of the design domain thus shifting the bottom boundary. By doing so, elements on the bottom edge get converted from boundary elements to interior elements. These dummy elements have no design variables associated with them but participate in objective and constraint function evaluations. Justification behind the special boundary consideration is discussed in section \ref{sec:discussion}.

\subsubsection{Displacement inverter problem}
\begin{figure}[h]
	\centering
	\subcaptionbox{Solution for mesh size: $60\times 30,~ vf = 0.22,~ ls = 1~\text{and}~ \lambda = 10^5.~ g(\boldsymbol{\rho}) = 2.7\times 10^{-3} $ }{\includegraphics[width=0.45\textwidth]{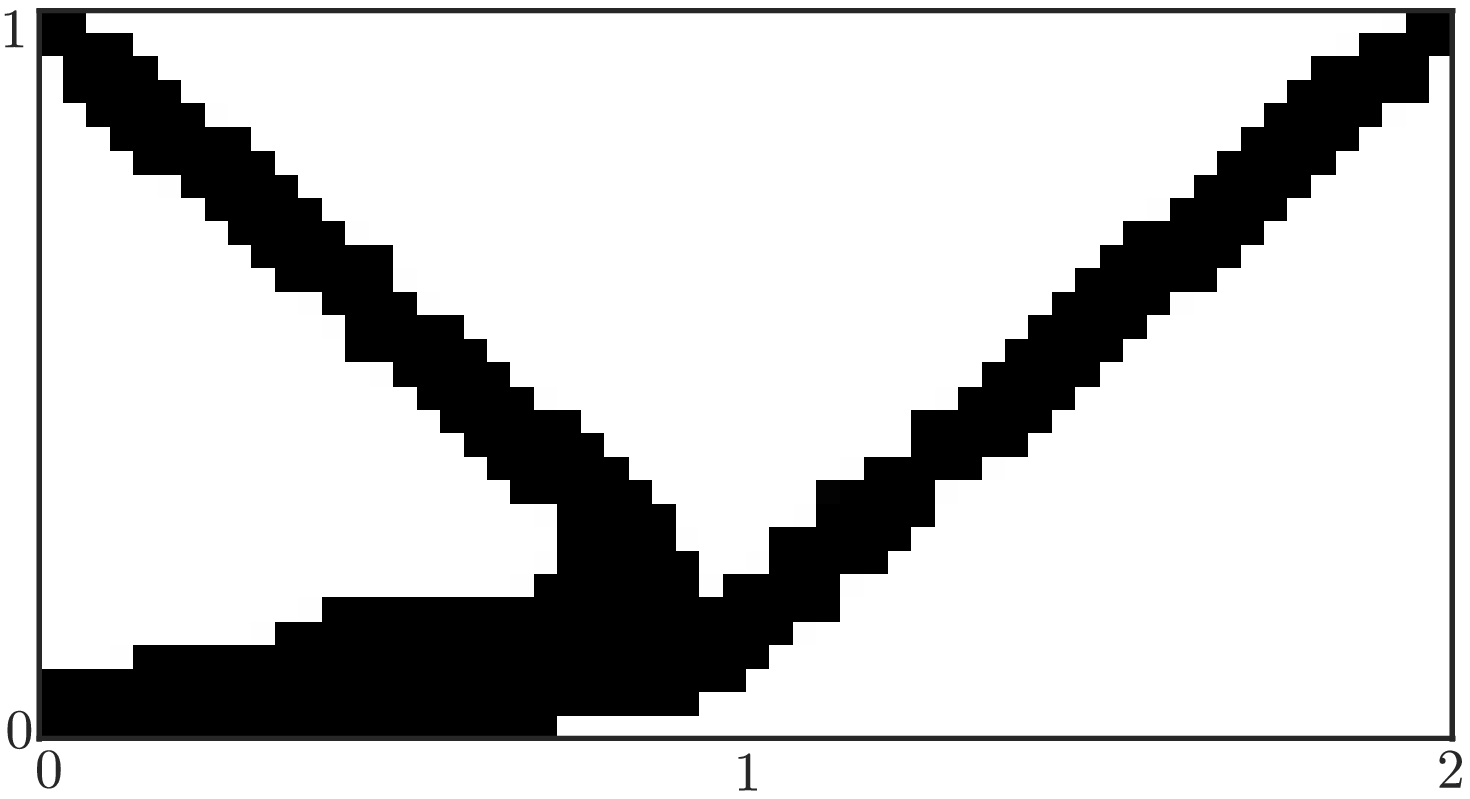}}
	\label{fig:Sol_Disp_inverter_60x30_vf_22_ls_1}%
	\hspace{0.4cm}
	%	\\[12pt]
	\subcaptionbox{Solution for mesh size: $100\times 50,~ vf = 0.22,~ ls = 2~\text{and}~ \lambda = 2\times 10^5.~ g(\boldsymbol{\rho}) = 4.1 \times 10^{-3} $}{\includegraphics[width=0.45\textwidth]{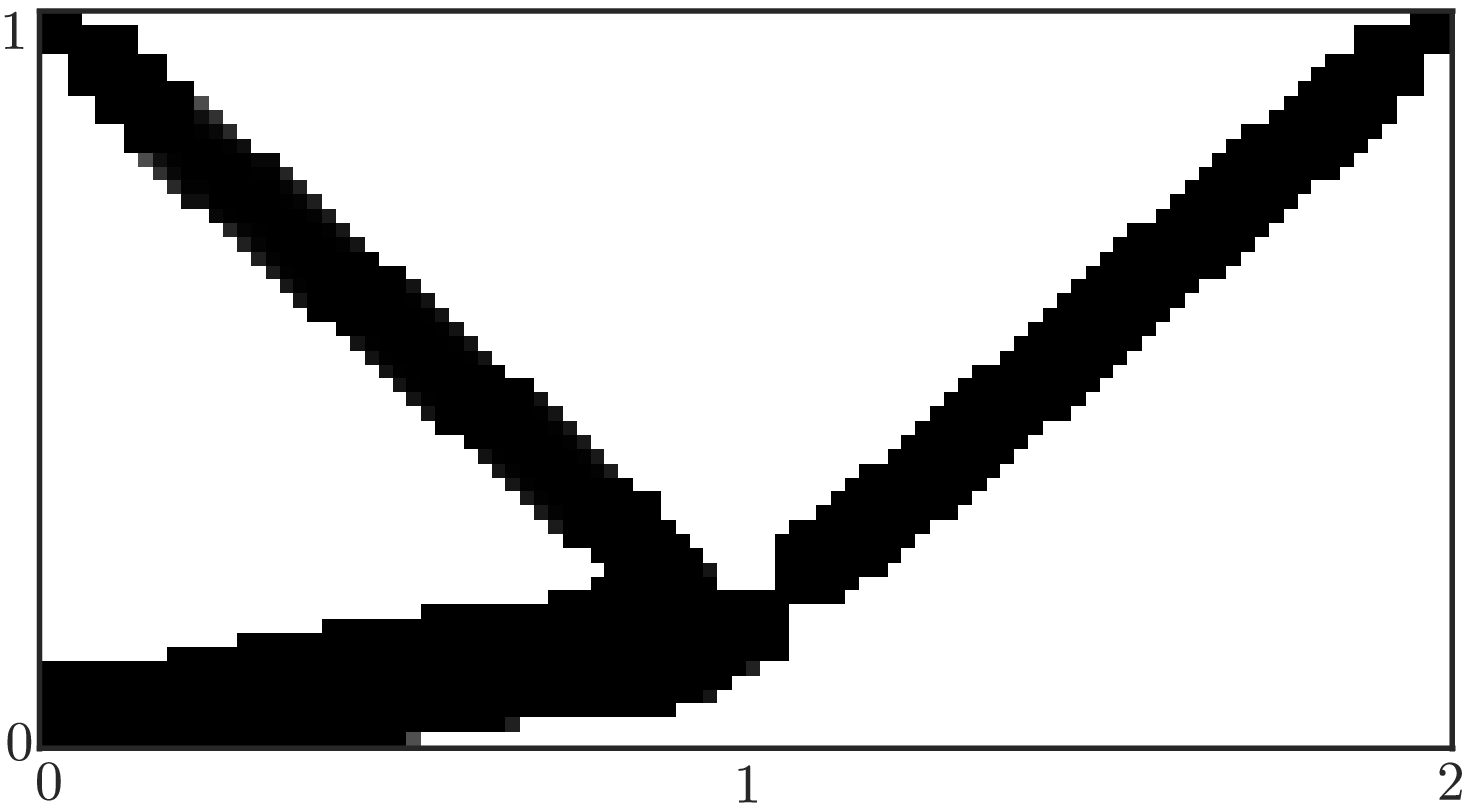}}
	\label{fig:Sol_Disp_inverter_100x50_vf_22_ls_2}%
	%	\hspace{0.7cm}
	\\[12pt]
	\subcaptionbox{Solution for mesh size: $140\times 70,~ vf = 0.22,~ ls = 3~\text{and}~ \lambda = 5\times 10^5.~ g(\boldsymbol{\rho}) = 1.13 \times 10^{-2} $}{\includegraphics[width=0.45\textwidth]{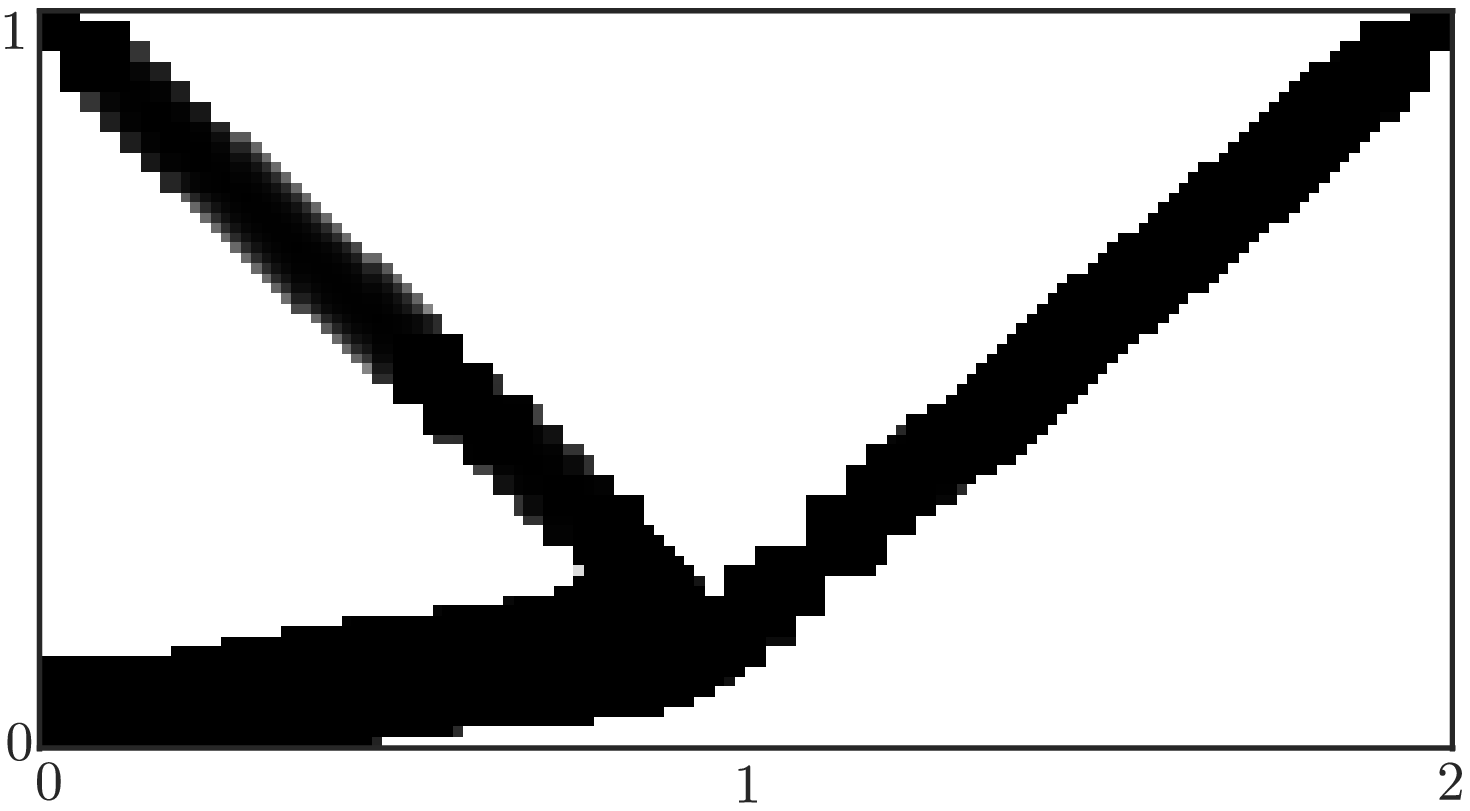}}
	\label{fig:Sol_Disp_inverter_140x70_vf_22_ls_3}%
	\hspace{0.4cm} 
	%	\\[12pt]
	\subcaptionbox{Solution for mesh size: $120\times 60,~ vf = 0.20,~ ls = 1~\text{and}~ \lambda = 10^5.~ g(\boldsymbol{\rho}) = 2.0 \times 10^{-3} $}{\includegraphics[width=0.45\textwidth]{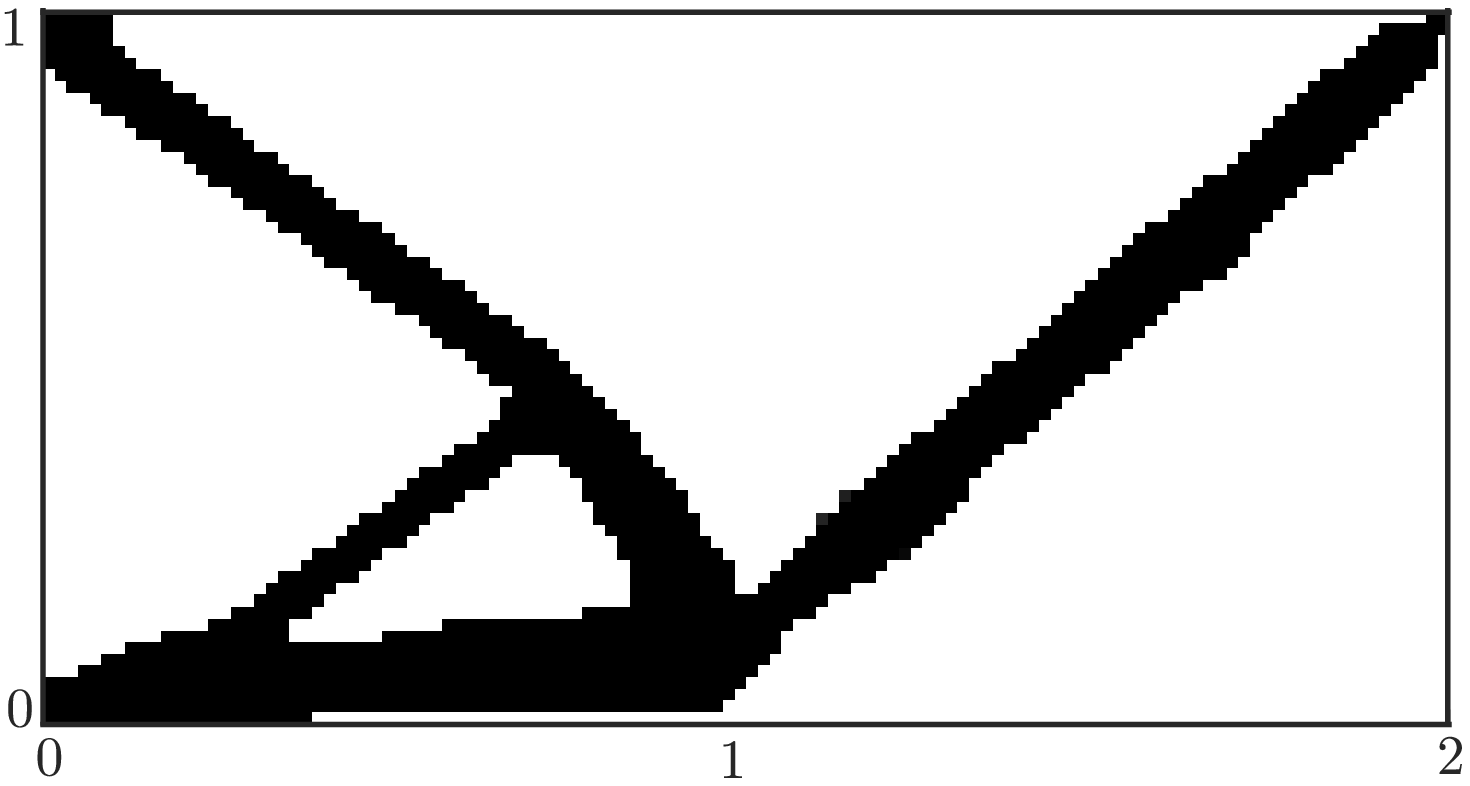}}
	\label{fig:Sol_Disp_inverter_120x60_vf_20_ls_1}%
	\\[12pt]
	\subcaptionbox{Solution for mesh size: $120\times 60,~ vf = 0.20,~ ls = 2~\text{and}~ \lambda = 10^5.~ g(\boldsymbol{\rho}) = 1.6 \times 10^{-3} $}{\includegraphics[width=0.45\textwidth]{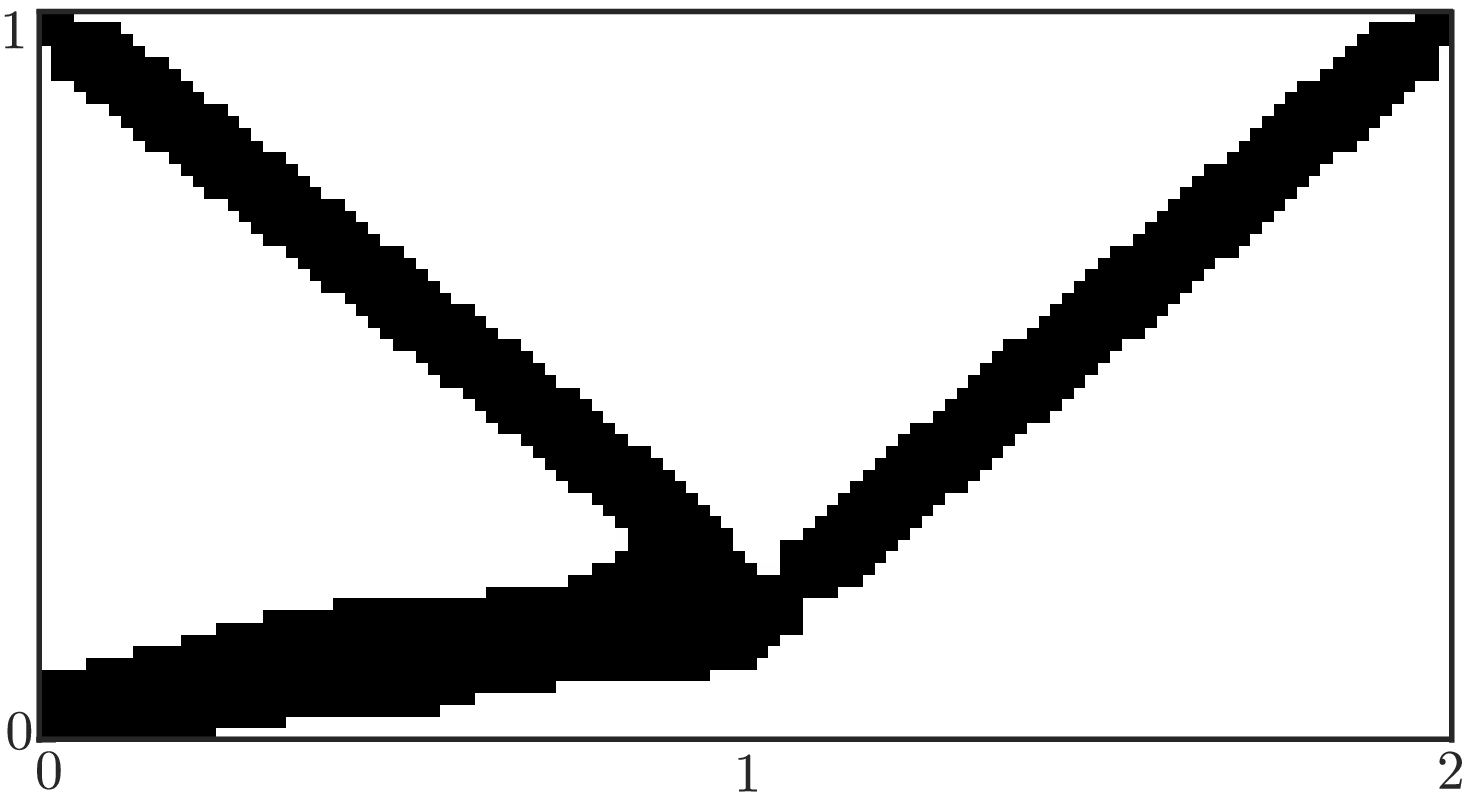}}
	\label{fig:Sol_Disp_inverter_120x60_vf_20_ls_2}
	\hspace{0.4cm}
	\subcaptionbox{Solution for mesh size: $120\times 60,~ vf = 0.30,~ ls = 2~\text{and}~ \lambda = 10^5.~ g(\boldsymbol{\rho}) = 2.2 \times 10^{-3} $}{\includegraphics[width=0.45\textwidth]{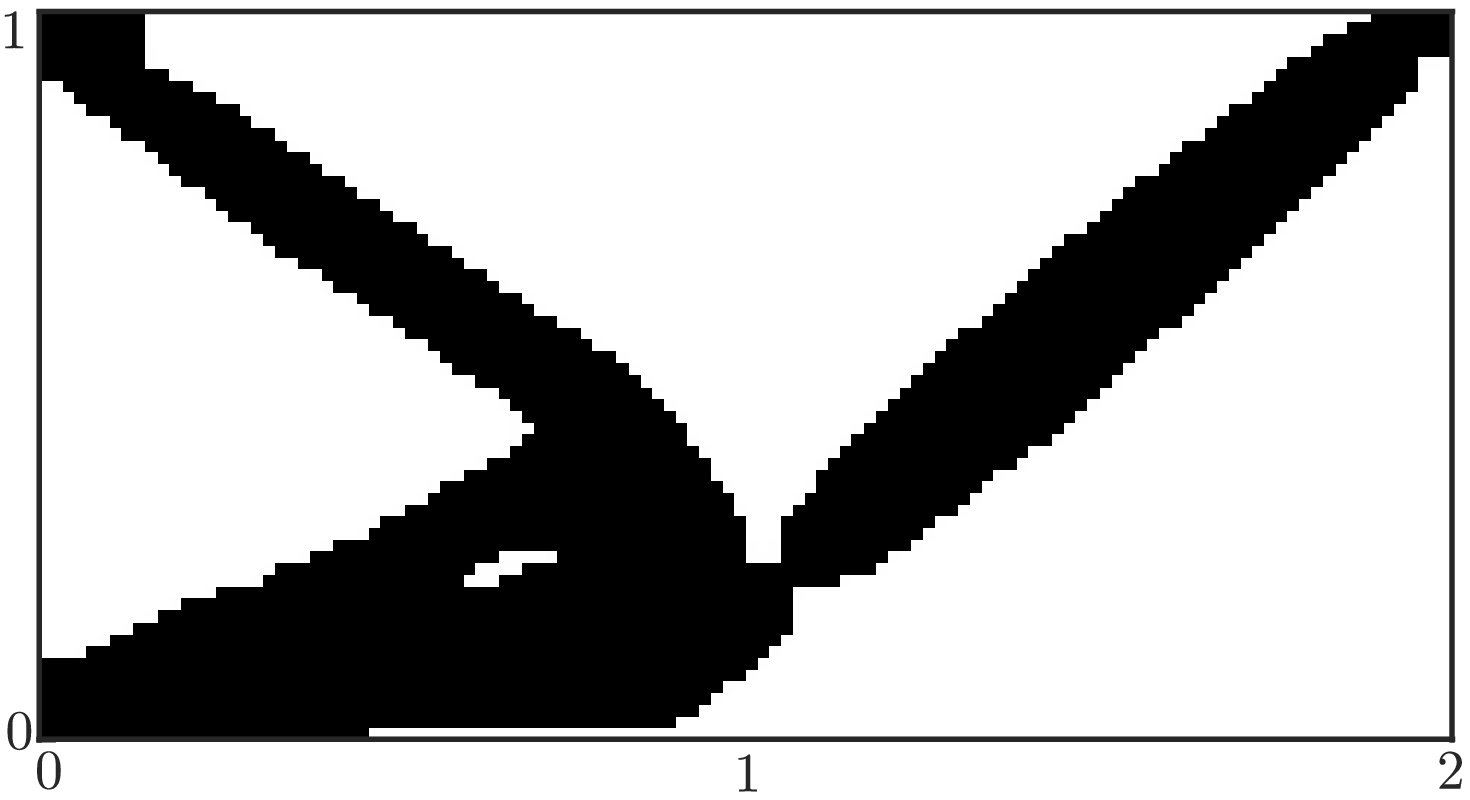}}
	\label{fig:Sol_Disp_inverter_120x60_vf_30_ls_3}
	\\[12pt]
	\subcaptionbox{Convergence history for solution presented in fig. \ref{fig:Sol_Disp_inverter}f }{\includegraphics[width=0.45\textwidth]{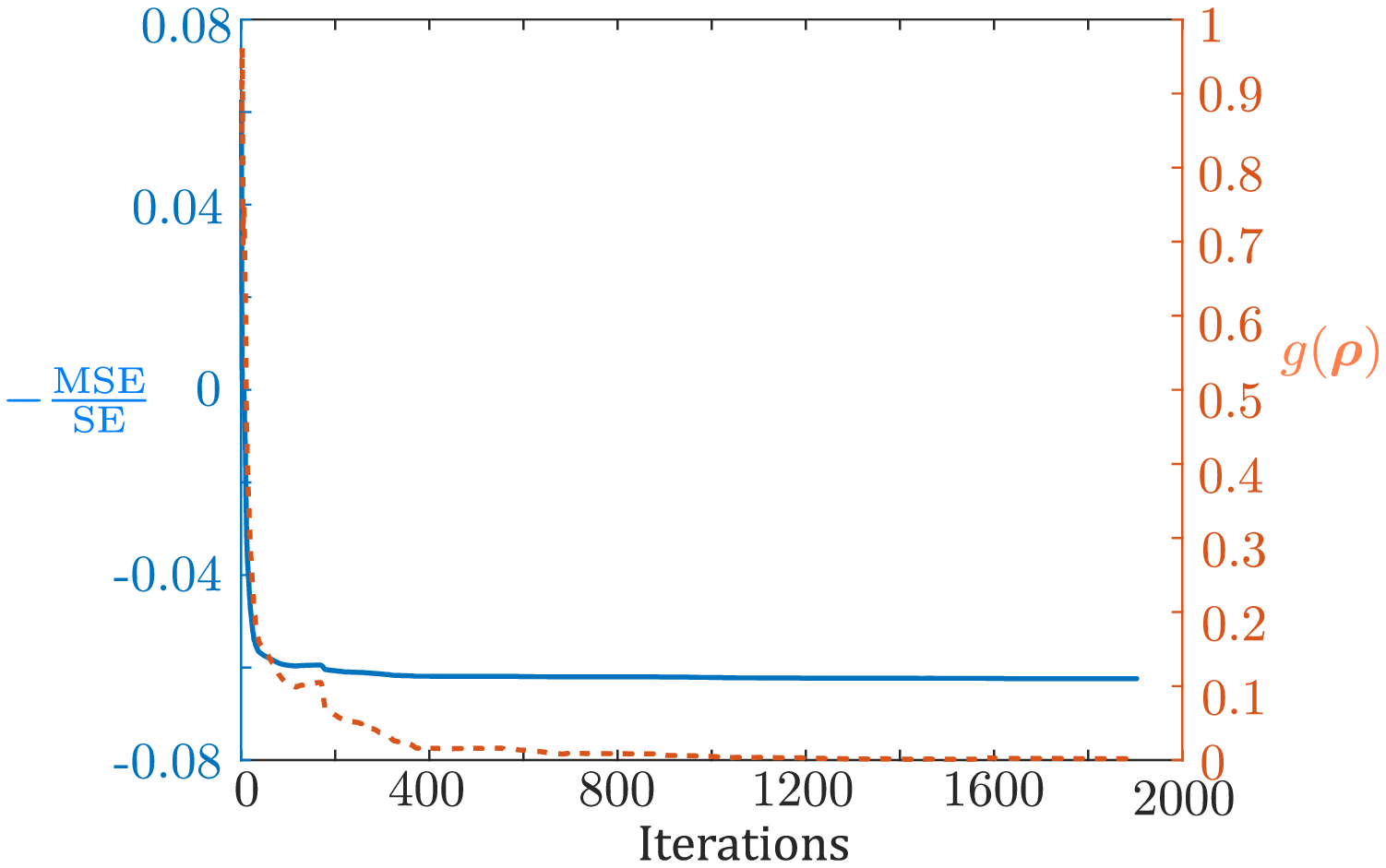}}
	\label{fig:Sol_Disp_inverter_120x60_vf_30_ls_3_converge}
	\caption{Solutions to the displacement inverter problem.}
	\label{fig:Sol_Disp_inverter}
\end{figure}

Fig. \ref{fig:Sol_Disp_inverter} presents solution to the displacement inverter problem for six different cases. We implement the same boundary consideration as in MBB problem to obtain these solutions. Domain is discretized into $60 \times 30$, $100 \times 50$ and $140 \times 70$ elements respectively in Figs. \ref{fig:Sol_Disp_inverter}a-c. Length scale $ls$ of 1, 2 and 3 respectively is implemented. Figures present topologies for varied domain discretization while maintaining the shape and size of $\Gamma(\boldsymbol{\mathrm{X}})$. Volume fraction for these solutions is $vf = 0.20$. Fig. \ref{fig:Sol_Disp_inverter}d, e and f present solutions on a $120 \times 60$ mesh for $vf = 0.20$ and $ls = 1$, $vf = 0.20$ and $ls = 2$, and $vf = 0.30$ and $ls = 3$ respectively. Solutions in fig. \ref{fig:Sol_Disp_inverter}a-c exhibit identical topologies showing mesh independence of the solution. Fig. \ref{fig:Sol_Disp_inverter}d-f present varied topologies. 
%The $g(\boldsymbol{\rho})$ for each solution is mentioned in captions below the figure, along with various other parameters implemented to obtain the solution. 
$g(\boldsymbol{\rho})$ for all solutions is close to or below $10^{-2}$ or $1\%$. For fig. \ref{fig:Sol_Disp_inverter}c gray cells are found at the top left corner of the structure. Presence of there gray cells is discussed in section \ref{sec:discussion}. Fig. \ref{fig:Sol_Disp_inverter}g presents the convergence history for the solution in fig. \ref{fig:Sol_Disp_inverter}f. Similar to previous examples, $g(\boldsymbol{\rho})$ takes much longer to converge and drop below an acceptable limit. 

%	\pagebreak
%	
	\section{Discussion}
\label{sec:discussion}
The normalized Field Product (nFP) method, proposed herein, yields mesh independent, pure black and white solutions without the need for any parameter or function choices involved in well connected density field evaluation, thus being free from any heuristic. The proposed method satisfies all criteria for an ideal density evaluation method mentioned in section \ref{sec:formulation_def}. Unlike projection \citep{Guest2004} or morphological filters \citep{sigmund2007morphology} wherein the densities are in some way forced towards 0-1 values, with the nFP method, the solutions gravitate naturally, without any explicit effort, towards pure 0-1 topologies. 
%$g(\boldsymbol{\rho})$ of the order of $10^{-3}$, along with the fact that pure 0-1 topologies are extremas for both the compliance minimization and compliant mechanism problem hint at the possibility that a pure 0-1 solution might be a local minima. 
%
%\nikcom{Note that a topology with $\rho(\boldsymbol{\mathrm{X}}) = 0$ or $1$ is not possible. Eqn. \ref{eqn:FPM_analy} enforces that $\rho(\boldsymbol{\mathrm{X}})$ is continuous even for a discontinuous $\beta(\boldsymbol{\mathrm{X}})$. Thus for the analytical field $\rho(\boldsymbol{\mathrm{X}})$ there will always exist a transition region at structural boundaries. A corollary of the discussions in section \ref{sec:conv_zero} is that these transition regions can be infinitesimally small. This is not possible in projection or filtering by design. The weighted average implemented in those methods will always take a finite transition region.}
%\textit{$\rho_i = 1$ is not part of the solution space and this the potential local minima at 0-1 topologies cannot be reached.}
%
%

As stated in section \ref{sec:results}, some results present gray cells at structural boundaries. This is usually a consequence of slow convergence rate. Convergence histories make it clear that the decline in $g(\boldsymbol{\rho})$ is gradual, hence, eliminating all gray cells will require a large number of iterations. In almost all cases, the method yielded the final solution giving $g(\boldsymbol{\rho})$ within an acceptable limit but with gray cells at few locations of the structural boundary. While in some cases, gradient of the objective with respect to cell densities of gray cells were too small, and computational local minima was achieved while gray cells were present in the topology. A \textit{monotonic decline} in $g(\boldsymbol{\rho})$ with number of iterations is observed for all problems even though other than the application of SIMP material model, no measures were taken to reduce/influence the grayness measure. Even though all solutions presented in section \ref{sec:example} are very close to 0-1 topologies, the formulation does not always lead to such solutions. Numerical experiments revealed that depending on the initial guess, optimization can converge to undesirable local minima while, in some rare cases, the gradient magnitude for a gray topology can be very small resulting in gray solutions (fig. \ref{fig:Sol_Disp_inverter}c). Thus, the method does not guarantee a 0-1 solution.

Definition of neighborhood and its discretization plays an important role in topology optimization. Similar to solid phase projection and filtering, the definition of neighborhood imposes an implicit length scale constraint on the solid phase. Thus, varying the definition of neighborhood alters the minimum member size, and hence changes are observed in the solution topology (fig. \ref{fig:Sol_Canti}, \ref{fig:Sol_Disp_inverter}). As with any other method for implicit imposition of length scales, compliance minimization solutions are well behaved while local violations are observed for the compliant mechanism problem. The nFP method imposes implicit length scale only on the solid phase and not the void phase. This is made evident by the solution in fig. \ref{fig:Sol_Disp_inverter}f, which displays a very slender void. Minor variations to the formulation can be made to impose length scale to the void phase instead to the solid phase. In addition to geometry, the discretization of neighborhood also affects the optimization process. Impact of discretization is reflected in eqn. \ref{eqn:density_grad}. As finer discretization is implemented, the ratio $\frac{A(\Omega_j)}{A(\Delta_i)}$ becomes smaller, reducing the magnitude of $\frac{\partial \rho_i}{\partial \beta_j}$. This in turn reduces the overall gradient magnitude which adversely affects the optimization process and may lead to gray solutions. Note that $\frac{A(\Omega_j)}{A(\Delta_i)}$ is independent of mesh size and instead depends on the length scale, $ls$. 

Another important aspect of implementing the nFP method is domain boundary consideration. The definition of neighborhood in section \ref{sec:Boundary_consideration} for boundary elements allows for thinner members at the boundary, giving boundary elements an advantage over interior elements. This can be seen in solutions to the cantilever beam problem where thin members are present along the bottom edge of the domain. Implementing the same approach hinders the optimization process in MBB beam and displacement inverter design problems especially for low volume fractions. Thus, special boundary consideration is implemented for these problems as discussed in section \ref{sec:results}. \cite{Guest2004} discusses three different approaches to deal with boundary elements when working with projection method. In an ideal case, one would want the boundary consideration to not hinder with the optimization irrespective of the problem and prevent thin members along the domain boundary.

As stated earlier, in the nFP method, solutions naturally gravitate towards 0-1 topologies giving low $g(\boldsymbol{\rho})$. The reason for this phenomenon is unclear at this time. Solutions with $g(\boldsymbol{\rho})$ below $1\%$ along with the fact that gradient magnitudes diminish as one approaches binary topologies hint at the possibility that local minima lie at upper and lower bound of design variables. Further mathematical investigation is required to better understand these observations.

An evident drawback of the nFP method, as with any other density based method, is that the number of design variables equals the number of finite elements. Thus, number of design variables increase with mesh refinement. This is a point of concern for 3-dimensional problems with large meshes. As the nFP method, in many cases, gives no transition region irrespective of the mesh, a limitation of the method is that it can only take spaces offered by the mesh, that is, a rectangular mesh such as one used herein cannot generate perfectly meshed slanted edges while a triangular mesh can have that possibility. This can lead to notches and points of stress concentration. Therefore, it is recommended that the final solution be passed through a shape optimization or a boundary smoothening process. Hexagonal tessellation \citep{saxena2003honeycomb, saxena2007honeycomb} can also be implemented to achieve smoother boundaries and remove the possibility of point connections. Numerical experiments reveal that the method works best with gray initial guesses. This is a consequence of gradient magnitudes diminishing in regions with densities close to 0 or 1.
%
%\begin{enumerate}
%	\item Black and white solutions are achieved without any parameter or function choices.
%	\item Gray solutions are allowed but the solutions gravitate towards black and white.
%	\item Gray cells are present in the topology at some places most likely due to small magnitude of gradients.
%	\item The method does not always lead to B/W solution.
%\end{enumerate}
%	\pagebreak
%
	\section{Conclusion}
\label{sec:conclusion}
The paper presents a novel, parameter free, density evaluation method for topology optimization based on normalized product of a scalar field over a domain. The approach allows for pure 0-1 solutions irrespective of the mesh refinement, imposes an implicit length scale on the solid phase and does not rely on any user based parameters. We couple the proposed density evaluation method with the SIMP material model. 

Solutions to two compliance minimization and one compliant mechanism problem are presented for various mesh refinements, volume fractions and length scales. The results obtained are close to pure 0-1, giving grayness measure below $1\%$ in most cases. By presenting solutions of fixed physical length scale and volume fraction for different mesh refinements, we establish mesh independence for both compliance minimization and compliant mechanism problems. The solutions also satisfy the length scale criterion throughout for compliance minimization problems while local violations of the same were observed in the flexibility-stiffness compliant mechanism solutions, as expected. Further, convergence history reveals that obtaining close to 0-1 solutions required a large number of iterations. This is a consequence of small gradient magnitudes in case of close to 0-1 topologies.
The methodology proposed looks promising and can be extended to various topology optimization problems. From an application point of view, the proposed method uses a setup similar to the projection method, while there is considerable difference in the expression implemented to evaluate element densities.

	\bibliography{opti_ref_1}
	
	\appendix
\section{Appendix A}
%\section{}
\label{Appendix:A}
Here we describe the method that can be implemented to determine $\boldsymbol{\beta}$ for a given density distribution, $\boldsymbol{\rho}$. Note that this is not required for implementing the nFP method but is discussed for the sake of completeness. From eqn. \ref{eqn:FPM_num}, 
\begin{eqnarray}\nonumber
&~&\rho_i = 1 - \exp \left(\dfrac{\sum\limits_{j \in \{\mathbb{D}_i\}}\beta_j A(\Omega_j)}{A(\Delta_i)}\right)\\[8pt]
\implies &~& A(\Delta_i) \ln(1- \rho_i) = \sum\limits_{j \in \{\mathbb{D}_i\}}\beta_j A(\Omega_j)
\end{eqnarray}
where $\beta_j \leq 0$. For the special case discussed in this work, area of all elements is the same, that is, $A(\Omega_j) = a$ and thus, the above equation can be simplified as,
\begin{flalign}\nonumber
&~& \sum\limits_{j \in \{\mathbb{D}_i\}}\beta_j &= \frac{A(\Delta_i) }{a}\ln(1- \rho_i) &\\[8pt] \label{eqn:lin_beta}
&~& \boldsymbol{\mathrm{C}} \boldsymbol{\beta} &=  \boldsymbol{\mathrm{b}} &\\[8pt]\nonumber
& \text{where} & \mathrm{C}(i,j) &=\begin{cases*}
 1 \text{~~~for~~~} j \in \{\mathbb{D}_i\}\\
 0 \text{~~~otherwise}
\end{cases*}
&\\[8pt]\nonumber
&\text{and}& b_i &= \frac{A(\Delta_i) }{a}\ln(1- \rho_i).&
\end{flalign}
Solving the linear system of equations in eqn. \ref{eqn:lin_beta} provides the value for $\boldsymbol{\beta}$. The system can only be solved for an invertible $\boldsymbol{\mathrm{C}}$. $\boldsymbol{\mathrm{C}}$ will be singular if there exists a $\{\mathbb{D}_i\}$ which is a union of two or more $\{\mathbb{D}_j\}$. The method implemented to define neighborhood in this work automatically ensures that $\boldsymbol{\mathrm{C}}$ is invertible.

Note that eqn. \ref{eqn:FPM_num} does not allow for all possible density distributions while eqn. \ref{eqn:lin_beta} can be solved for any density distribution. Solving for density distribution which cannot be achieved by eqn. \ref{eqn:FPM_num} will lead to positive values of $\beta_i$ which is not allowed in nFP. 

The evaluation method presented above even though analytically sound, renders moot in the numerical setting. This is because as any of the elemental densities converges to $1$ the evaluation of $\ln(1-\rho_i)$ goes beyond the scope of numerical evaluation and thus the corresponding $\boldsymbol{\beta}$ cannot be evaluated accurately.

\end{document}